\documentclass[
twocolumn,
]{aastex631}

\usepackage{amsmath}
\usepackage{xcolor}
\usepackage{appendix}
\usepackage{array}
\usepackage{hyperref}

\newcommand\thesis[1]{}

\def\ie{{\it i.e.,}}
\def\eg{{\it e.g.,}}
\newcommand{\abcsn}{\texttt{ABC-SN}}
\newcommand{\lsst}{\textit{LSST}}
\newcommand{\snid}{\texttt{SNID}}
\newcommand{\dash}{\texttt{DASH}}
\newcommand{\sniascore}{\texttt{SNIascore}}
\newcommand{\sedmachine}{\textit{SEDMachine}}
\newcommand{\sedm}{\textit{SEDM}}
\newcommand{\ccsnscore}{\texttt{CCSNscore}}
\newcommand{\superfit}{\texttt{SUPERFIT}}
\newcommand{\gelato}{\texttt{GELATO}}
\newcommand{\ngsf}{\texttt{NGSF}}
\newcommand{\vgg}{\texttt{VGG-16}}

\newcommand{\nSpec}{$3,764$}
\newcommand{\nSN}{$498$}

\shorttitle{ABC-SN}
\shortauthors{Fortino et al.}

\graphicspath{{./}{}}

\begin{document}

\title{ABC-SN: Attention Based Classifier for Supernova Spectra}

\correspondingauthor{Willow Fox Fortino}

\author[0000-0001-7559-7890]{Willow Fox Fortino}\email{fortino@udel.edu}

\affiliation{Department of Physics and Astronomy, University of Delaware, Newark, DE 19716, USA}

\author[0000-0003-1953-8727]{Federica B. Bianco}

\affiliation{Department of Physics and Astronomy, University of Delaware, Newark, DE 19716, USA}
\affiliation{Joseph R. Biden, Jr. School of Public Policy and Administration, University of Delaware, Newark, DE 19716, USA}
\affiliation{Data Science Institute, University of Delaware, Newark, DE 19716, USA}
\affiliation{Vera C. Rubin Observatory, Tucson, AZ, USA}

\author[0000-0002-8178-8463]{Pavlos Protopapas}

\affiliation{John A. Paulson School of Engineering and Applied Sciences, Harvard University, Cambridge, MA, 02138, USA}

\author[0000-0002-5788-9280]{Daniel Muthukrishna}

\affiliation{Kavli Institute for Astrophysics and Space Research, Massachusetts Institute of Technology, Cambridge, MA 02139, USA}

\author[0000-0002-7293-8140]{Austin Brockmeier}

\affiliation{Data Science Institute, University of Delaware, Newark, DE 19716, USA}
\affiliation{Department of Electrical and Computer Engineering,
University of Delaware, Newark, DE, USA}
\affiliation{Department of Computer and Information Sciences, University of Delaware, Newark, DE, USA}

\begin{abstract}
    While significant advances have been made in photometric classification ahead of the millions of transient events and hundreds of supernovae (SNe) each night that the Vera C. Rubin Observatory Legacy Survey of Space and Time (\lsst{}) will discover, classifying SNe spectroscopically remains the best way to determine most subtypes of SNe. Traditional spectrum classification tools use template matching techniques \citep{blondin_determining_2007} and require significant human supervision. Two deep learning spectral classifiers, \dash{} \citep{muthukrishna_dash_2019} and \sniascore{} \citep{fremling_sniascore_2021} define the state of the art, but \sniascore{} is a binary classifier devoted to maximizing the purity of the SN Ia-norm sample, while \dash{} is no longer maintained and the original work suffers from contamination of multi-epoch spectra in the training and test sets. We have explored several neural network architectures in order to create a new automated method for classifying SN subtypes, settling on an attention-based model we call \abcsn{}. We benchmark our results against an updated version of \dash{}, thus providing the community with an up-to-date general-purpose SN classifier. Our dataset is comprised of ten different SN subtypes including subtypes of SN Ia, core collapse and interacting SNe. We find that \abcsn{} outperforms \dash{}, for nearly all classes, including an improvement of 26\% in SN Ia completeness ($\sim88$\%) and 2.4\% in SN Ia purity ($\sim95$\%) when unthresholded (improvements for each class can further be obtained by tuned thresholds) and we discuss the limitation of current SN datasets for benchmarking performance.
\end{abstract}

\section{Introduction} \label{sec:intro}
As the field of astronomy stands on the brink of a data revolution brought about by the Vera C. Rubin Observatory's Legacy Survey of Space and Time (\lsst{}), the astronomy community braces for an influx of data at an unprecedented scale. The \lsst{} will observe millions of transient events nightly, with supernovae (SNe) poised to form a substantial fraction. This will stretch the existing spectral follow-up infrastructure to its limits, necessitating a paradigm shift in our approach to handling the flood of SN candidates that will emerge from the \lsst{} alert stream. Effective classification of SN subtypes is crucial for our understanding of stellar evolution and the distribution of elements in the universe.

\citet{minkowski_spectra_1941} introduced the first SN classification scheme by dividing the SNe that did not contain hydrogen (Type I) from those that did (Type II). In his review, \citet{filippenko_optical_1997} describes the evolved classification scheme as of the late 1990s, with further subclasses of SNe based on the early-epoch chemical composition identified in the spectra. While all Type I lack the presence of hydrogen (H) lines, Type Ia spectra display intense Si\textsc{ii}\footnote{Singly-ionized Si.} lines, Type Ib spectra exhibit pronounced He\textsc{i} lines, and Type Ic spectra show neither (see also \citealt{2016ApJ...820...75P, 2001AJ....121.1648M, williamson2023sn}).

SNe Ib, Ic, and II are core-collapse SNe \citep{janka2012core, 2019NatAs...3..717M} with progenitor stars of a mass $> 8 M_\odot$  \thesis{that end their lives after the stellar core fuses iron (Fe), fusion becomes energetically inefficient, and gravity is no longer balanced by radiation pressure } \citep{woosley1986physics, woosley2005physics, couch_mechanisms_2017}. On the other hand, SNe Ia are the result of thermonuclear explosions \citep{ropke2007thermonuclear} produced when a white dwarf (WD) star is accreting matter from a nearby star \citep{mazzali_common_2007} beyond the ``Chandrasekhar limit'' of $\sim 1.4 M_\odot$ (although in practice both sub- and super-Chandrasekhar progenitors have been identified, \eg{} \citealt{andrew2006type, hicken2007luminous}, see for example \citealt{woosley2011sub} for a theoretical approach). Carbon fusion occurs quickly and explosively, producing the SNe that are observed. The intrinsic brightness of SNe Ia can be calculated from their temporal evolution and color, and they are therefore known as ``standard(izable) candles''. Cosmologists use SNe Ia to study dark energy and the expansion of the Universe \citep{riess_observational_1998, perlmutter_measurements_1999}.

Yet, further `subtypes' of the mentioned SN types exist that are also distinguished by their spectral lines. Some progenitors are surrounded by circumstellar material (CSM), which interacts with their SNe, causing differences in spectra and leading to their own classification (e.g., SN IIn and Ibn). Some SNe are identified as abnormal and receive their own classification. Particularly important for cosmology is the identification of specific SN Ia subtypes (e.g., SN Ia 91T, 91bg) whose properties may deviate from the canonically formalized standardization that enables measures of the Hubble constant \citep[\eg][]{2016MNRAS.461.2044S, blondin_spectroscopic_2012}. In some cases, photometric properties also modify a classification, as in the case of SN IIP, showing a plateau in the light curve and SN IIL, showing a linear decrease while, ostensibly, they may present the same spectroscopically (\citealt{2017hsn..book..195G}, but also see \citealt{2020ApJ...890..177K}, conversely, note that spectroscopically unusual SNe may present within the normal range photometrically \citealt{khakpash2024multifilter}).

Additional explosion methods \citep[\eg{} pair instability][]{woosley2017pulsational, kasen2011pair}, and several more types and subtypes of Super-Luminous Supernovae exist, but are not addressed in this work; see \citet{moriya2018superluminous} for a review.

Altogether, this brief introduction only scratches the surface of a complex, and apparently ever-growing taxonomy. Typing SNe can involve completely different progenitors or explosion mechanisms along with nuanced differences in continuous parameters: how much material was stripped before the explosion, progenitor mass, metallicity of the progenitor environment, etc. The reader should keep this in mind and throughout this report we discuss how these complications may impact the results of automated classifiers.

While photometric classification methods \citep[\eg{}][]{ 2019AJ....158..257B, 2020MNRAS.491.4277M, 2021AJ....162...67Q, 2025arXiv250101496S} have made significant advances motivated by the upcoming \lsst{} which will discover too many and too faint SNe for comprehensive spectroscopic classification, and with multi-modal models starting to appear \citep{2024MLS&T...5d5069Z}, classifying SNe spectroscopically remains the only way to determine most subtypes of SNe. Photometric classifiers can reach high accuracy in separating broad classes, but none of the photometric classifier can differentiate the stripped envelope supernova subtypes \citep{khakpash2024multifilter}: SN IIb, Ib, Ic, Ic-broad, Ibn etc., which are usually classified together as ``SN Ibc''.
Currently, the most prevalent methods for the classification of SN subtypes are based on template matching: Supernova Identification \citep[][\snid{}]{blondin_determining_2007} and \superfit{} \citep[][recently automated in Python as Next Generation SuperFit \ngsf{}, \citealt{2022TNSAN.191....1G}]{howell_gemini_2005} or GEneric cLAssification TOol \citep[\gelato{}]{harutyunyan_esc_2008}. Finally, we note that multimodal foundation models are now appearing in the literature (e.g. \citealt{zhang2024maven}), a very promising upgrade from single modality classifiers, but still in early concept phases.

\snid{} types new SNe with an extensive library of labeled SNe and template spectra (we will use the \snid{} library as our dataset and updated \snid{} labels as our ground truth, see \autoref{sec:data}). Though this approach has been a mainstay in the field due to its reliability and interpretability, the oncoming data deluge from the \lsst{} presents a scalability challenge that \snid{} may struggle to meet efficiently. As SN science evolves, template matching methods require continuous, careful reclassification. Meanwhile, machine learning models like \sniascore{} \citep{fremling_sniascore_2021} and its sibling \ccsnscore{} \citep{sharma_ccsnscore_2025}, which have shown proficiency in classifying Type Ia SNe and separating Hydrogen-rich from Hydrogen-poor SNe with the low-resolution spectral data from the \sedmachine{} \citep[][\sedm{}]{blagorodnova_sed_2018}, have yet to demonstrate the same adaptability for a wider variety of SN subtypes and spectra sourced from different instruments.

This paper introduces \abcsn{} the `Attention-Based Classifier for Supernovae', a novel approach to SN classification that leverages the multi-head attention layer in the transformer neural network architecture \citep{vaswani_attention_2017}, a model that has significantly impacted the field of natural language processing and was subsequently adapted for innumerable applications from computer vision \citep{dosovitskiy2020image} to time series forecasting \citep{zhou2021informer}. Like the complex web of syntax and semantics in human languages, SN spectra contain intricate patterns --- spectral features --- that signal the presence of specific elements, indicate stellar structure patterns, and determine a SN's subtype. We posit that the transformer's ability to parse these complex patterns in language could be directly applicable to spectral classification. Applications of transformer-derived architectures to spectroscopy outside of astrophysics include \citep{cai2022mst++} and a number of models for application on hyperspectral imaging \citep[\eg][]{he2021spatial}. In astrophysics, see  for example \citet{2024MLS&T...5d5069Z} and \citet{koblischke_spectrafm_2024}.

An important precursor to our work is the Deep Automated Supernova and Host (\dash{}) classifier \citep{muthukrishna_dash_2019}. \dash{} is a convolutional neural network designed to determine subtype, age, redshift, and host galaxy from SN spectra. Compared to subtype classification, \dash{} tackles a significantly more complex task. Because \dash{} is the only other general-purpose, machine learning based, spectral SN classifier, we will compare \abcsn{} to \dash{}'s performance as a subtype classifier by aggregating over its age classification.

The goal of this research is to introduce a classifier that surpasses the limitations of \snid{}, \sniascore{}, and \dash{}. To that end, we have assembled a dataset of high-resolution SN spectra to train this new attention-based classifier on. The full composition, provenance, and preprocessing of the data are discussed in \autoref{sec:data}. The details of how \abcsn{} is trained and evaluated are discussed in \autoref{sec:methods}.

\begin{figure*}[t!]
    \includegraphics[width=\textwidth]{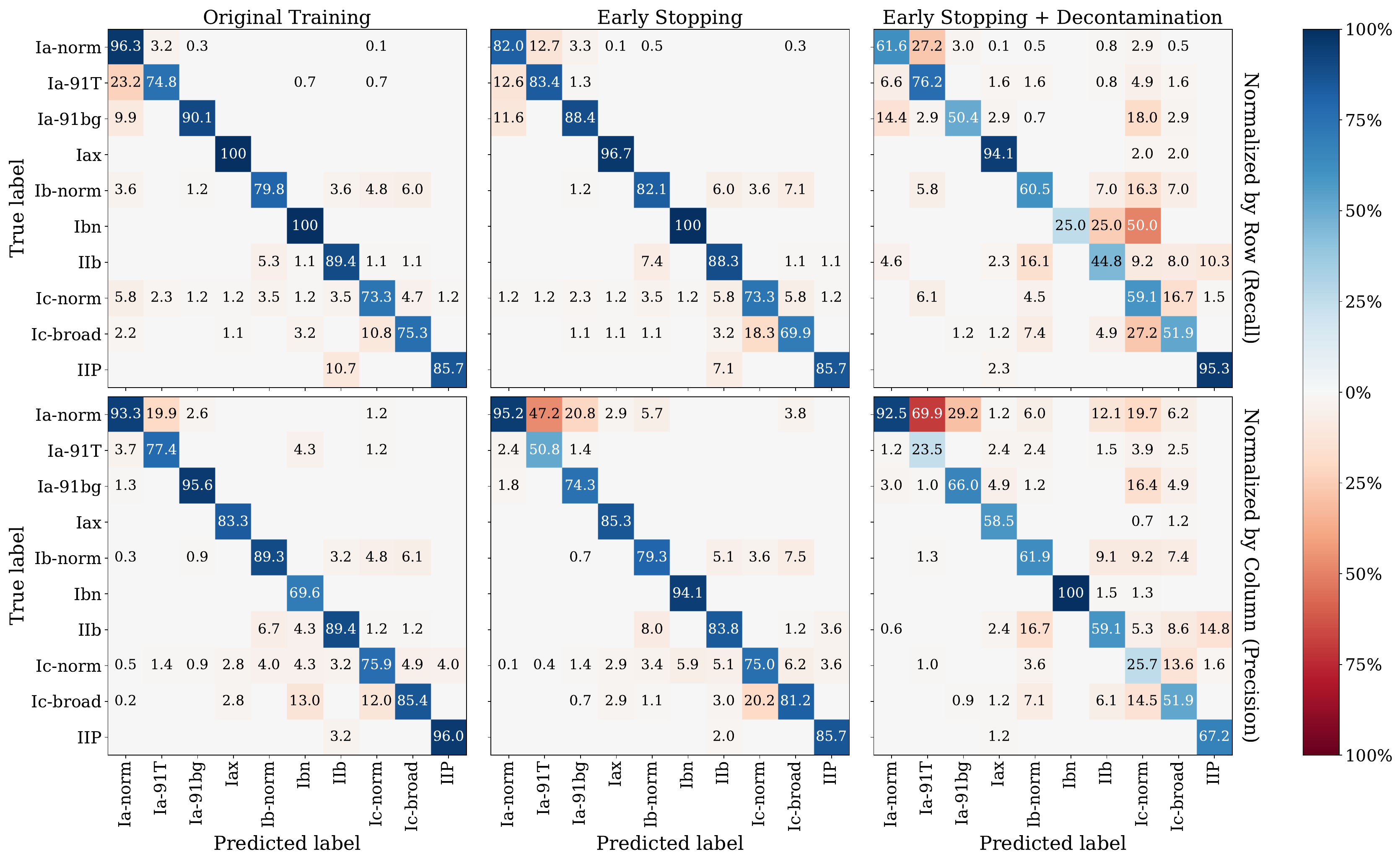}
    \centering
    \caption{\dash{} \citep{muthukrishna_dash_2019} SN classifier performance under different training schemas for the classes considered in this work (\autoref{sec:data:breakdown}). The {\it top} row shows model completeness (\ie{} recall) by normalizing across true labels. The {\it bottom} row shows purity (\ie{} precision) by normalizing across the predictions. In each confusion matrix, correct predictions (diagonal cells) are plotted in shades of blue  and incorrect ones (off-diagonal) in shades of red. In each cell, the percentage of spectra 
    is reported, but 0's are omitted. 
    {\it Left}: \dash{}  retrained with the original training schema, which includes the same SN at different epochs in the training and test set and allows training to complete to 12,500 pseudo-epochs (see \autoref{sec:retraindash}), inducing significant overfitting. While the confusion matrix looks best, the transferability and reliability of this model are in question due to severe overfitting. {\it Middle}: The same training and test sets are used, but training is stopped early to prevent overfitting (see \autoref{sec:retraindash}). {\it Right}: The train-test split has been modified to ensure that all spectra from each SN appear {\it only} in either the training or the test set. We will compare \abcsn{} to this final model. Note that \dash{} is retrained throughout on $R=100$ spectra (see \autoref{sec:data:prep:lowres}).}
    \label{fig:dash_compare}
\end{figure*}

\section{Retraining DASH} \label{sec:retraindash}
To establish a benchmark for the automated classification of SN spectra, we retrain \dash{} with four goals in mind: To replicate \dash{}'s published results; To determine \dash{}'s results when compensating for overfitting; To determine \dash{}'s results without dataset contamination; To provide a benchmark to compare \abcsn{} to. To that end, we retrained \dash{} in three different ways, the results of which are shown in \autoref{fig:dash_compare}.

In order to compare \abcsn{} to \dash{}, we must consider \dash{}'s training scheme. Using code from \citet{muthukrishna_dash_2019}, we set the training set fraction to 50\% and find that the initial $\sim 2,000$ spectra in both the training and test sets are magnified to $\sim 130,000$ spectra in each set after data augmentation. The details of the dataset, including the number of labels per type and subtype, and any modifications from the dataset originally used in \dash{} are discussed in \autoref{sec:data}. Here we focus on the training and hyperparameter choices made in retraining \dash. Yet, a word on notation is in order: due to the severe imbalance of our data, we cannot perform a train-validation-test split, nor can we apply a cross-validation strategy (and this would also deviate further from the approach taken in \citealt{muthukrishna_dash_2019}). We will return to this point in \autoref{sec:data:prep:culling_standardization}, and we will perform transferability tests in \autoref{sec:results}, but for now, the reader is advised that in the two-way split of our data we refer to the set used for learning parameters as \emph{training} and the set used for hyperparameter tuning as \emph{test}, in accord with the notation used, for example, in \citealt{bishop2006pattern}.\footnote{Although we note other authors refer to this as the \emph{validation} set, consistently with the set used for this purpose in a three-way training-validation-test split (\eg{} \citealt{bengio2017deep}).}

The batch size during \dash{} training is 50, which means that it takes $\sim 2600$ batch iterations to complete one training epoch. For this reason, we propose the concept of a `pseudo-epoch'. One pseudo-epoch is equivalent to the number of \dash{} batches it takes such that $50 \times N_{\rm batches} = N_{\rm train}$ where $N_{\rm train}$ is the size of the training set {\it before} data augmentation. The pseudo-epoch helps us compare training times and overfitting between \dash{} and \abcsn{} and does not affect the way that \dash{} is retrained.

A dataset of SN spectra poses a unique complication: the same SN may be observed multiple times at different epochs in the lifetime of the transient. For instance, our dataset is comprised of \nSpec{} spectra from \nSN{} SNe, where 144 SNe contribute 10 or more spectra each. While SNe do evolve, key signatures of the cosmic explosion will persist over time. Thus, all spectra from the same SN are highly correlated; they are typically more self-similar than similar to spectra of other SNe of the same subtype.

All spectra from one SN must go into either the training or the test set. Contamination of spectra from the same object in the training and test set, even if taken at different epochs, would artificially inflate the performance of the model once results were calculated on the test set.

However, since \dash{} was designed to identify the epoch at which a spectrum was collected (in addition to the SN type), this exact type of contamination affected \dash{}.

We retrain \dash{} in three steps (see \autoref{fig:dash_compare}). We first run \dash{} in a way that mimics the original published results: the contamination in the training and test sets is present, and the model is trained for $500,000$ batches. This satisfies our first goal (\autoref{fig:dash_compare}, left).

Then we curtail overfitting by implementing an early stopping protocol which halts training if model performance doesn't increase. Specifically, if test set F1-score does not increase by 0.005 for 10 pseudo-epochs, we stop training. Also, if the training set F1-score exceeds the test set by 0.10 for 10 pseudo-epochs, we halt training. This satisfies our second goal (\autoref{fig:dash_compare}, middle).

In the third and final run, we correct the dataset contamination while also implementing the early stopping protocol. This satisfies our third goal.

All three instances of \dash{} are trained with all original 16 SN subtypes. However, in \autoref{fig:dash_compare} we present confusion matrices for these models including only the ten subtypes seen in \autoref{fig:treemap}. We also note that we retrain \dash{} on a low-resolution ($R = \frac{\lambda}{\Delta \lambda} = 100$) spectral dataset (see \autoref{sec:data:prep:lowres} for justification). These two changes satisfy our fourth goal and thus provide a new performance benchmark for the automated classification of SN spectra with \dash{} (\autoref{fig:dash_compare}, right). This benchmark F1-score is $F1_\mathrm{DASH}=0.40$.

Note that \citet{fremling_sniascore_2021} also offers a comparison of \sniascore{} with \dash{}. However, they use low-resolution spectra from the \sedm{} and do not retrain \dash{} on low-resolution spectra. Thus, their performance comparison is not fair. We chose to retrain \dash{} on our dataset to offer a rigorous performance comparison.

In \autoref{sec:intro} we noted how \dash{} classifies the subtype \textit{and} age of spectra. Specifically, \dash{} outputs classification probabilities for 306 type-age classes (18 age classes for 17 subtypes). In order to compare \abcsn{} to \dash{} we aggregated all type-epoch classifications by type only. \footnote{For example, if \dash{} predicted that a spectrum was a SN Type Ia-norm at 2 to 6 days after peak, we would consider that simply a SN Type Ia-norm when comparing to \abcsn{} or calculating metrics.}

\section{Data} \label{sec:data}
\label{sec:data:breakdown}

\begin{figure}[t]
    \includegraphics[width=\columnwidth]{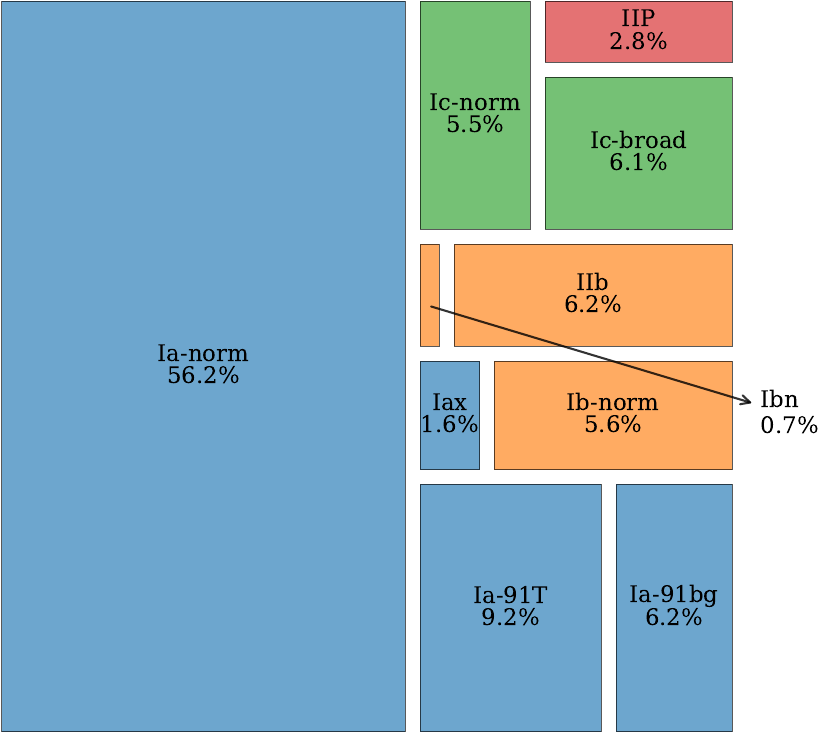}
    \centering
    \caption{A treemap plot showing the composition of our dataset of SN spectra after preprocessing. The size of each rectangle corresponds to the proportion of the ten SN subtypes in our dataset. The color of each rectangle corresponds to which main SN type --- Ia, Ib, Ic or II --- the subtype belongs to. There are \nSpec{} total spectra from \nSN{} SNe. This plot shows the extreme class imbalance in this dataset, with $\sim73\%$ of the spectra coming from SN Ia. This poses a profound challenge for machine learning. The content of this plot is equivalent to the content of the second column in \autoref{table:data}.}
    \label{fig:treemap}
\end{figure}

\begin{table*}[t!]
    \centering
    \begin{tabular}{l|lllll}
    SN Subtype (SNe) & Before PP & After PP & Trn Set & Trn Set (w/ Aug) & Tst Set \\
    \hline      
    Ia-norm (319) & 2387 (53.5\%) & 2114 (56.2\%) & 1058 (56.2\%) & 1058 (9.6\%) & 1056 (56.2\%) \\
    Ia-91T (36) & 398 (8.9\%) & 348 (9.2\%) & 163 (8.7\%) & 1141 (10.3\%) & 185 (9.8\%) \\
    Ia-91bg (42) & 264 (5.9\%) & 232 (6.2\%) & 101 (5.4\%) & 1111 (10.1\%) & 131 (7.0\%) \\
    Iax (6) & 68 (1.5\%) & 62 (1.6\%) & 28 (1.5\%) & 1064 (9.6\%) & 34 (1.8\%) \\
    \hline
    Ib-norm (22) & 270 (6.0\%) & 211 (5.6\%) & 99 (5.3\%) & 1089 (9.9\%) & 112 (6.0\%) \\
    Ibn (3) & 31 (0.7\%) & 27 (0.7\%) & 9 (0.5\%) & 1062 (9.6\%) & 18 (1.0\%) \\
    IIb (19) & 328 (7.3\%) & 232 (6.2\%) & 139 (7.4\%) & 1112 (10.1\%) & 93 (4.9\%) \\
    \hline
    Ic-norm (21) & 263 (5.9\%) & 206 (5.5\%) & 112 (5.9\%) & 1120 (10.2\%) & 94 (5.0\%) \\
    Ic-broad (24) & 279 (6.2\%) & 228 (6.1\%) & 117 (6.2\%) & 1170 (10.6\%) & 111 (5.9\%) \\
    \hline
    IIP (6) & 176 (3.9\%) & 104 (2.8\%) & 58 (3.1\%) & 1102 (10.0\%) & 46 (2.4\%) \\
    \hline
    \end{tabular}
    \caption{This table shows the exact distribution of the number of SN spectra across each of the ten SN subtypes. The number of SNe for each subtype is shown in parentheses next to the subtype name. The {\it first column} shows the distribution of the spectra we collected before any preprocessing. The {\it second column} contains the same information as in \autoref{fig:treemap} and shows the distribution after the preprocessing steps, which remove some known bad spectra. The {\it third} and {\it fifth columns} show the distributions of the training and test set split. The {\it fourth column} shows the final distribution of the training set after the data augmentation techniques (see \autoref{sec:data:prep:augmentation}) have been applied. Note how each class is represented approximately evenly in the training set after augmentation, while the test set preserves the original class imbalance.}
    \label{table:data}
\end{table*}

A direct comparison with \dash{} was a goal from the outset, so we start with the same dataset used to train \dash{}: the spectral library that comprises the \snid{} templates. Furthermore, we start with the \snid{} labels as our ground truth. However, we make several updates and modifications to this dataset. The original spectra in our dataset were gathered from a variety of sources. In addition to the core \snid{} library, they contain a catalog of stripped envelope SNe (SESNe) that is included from \citet{liu_supernova_2015, modjaz_optical_2014, modjaz_spectral_2016, liu_analyzing_2016} (which were integrated into the SNID library after \dash{} was trained). SESNe are SNe where the progenitor star's outer shell of hydrogen and helium have been removed prior to explosion and refer to the subtypes Ib-norm, IIb, Ic-norm and Ic-broad.

We also include a catalog of spectra from the Berkeley SN Ia Program \citep[][{\it BSNIP}]{silverman_berkeley_2012}.

The \snid{} templates are described in \citet{blondin_determining_2007, blondin_spectroscopic_2012} and were originally collected from the \textit{SUSPECT} public archive \citep{yaron_wiserepinteractive_2012}, the CfA Supernova Archive and the CfA Supernova Program \citep{matheson_optical_2008, blondin_spectroscopic_2012} and were retrieved by \citet{muthukrishna_dash_2019}.

Our dataset is composed of \nSpec{} spectra from \nSN{} SNe at a spectral resolution $R \equiv 738$. These spectra were selected because of their generally high signal-to-noise-ratio (SNR) and uniformity of preprocessing. The dataset represents 17 SN subtypes, but there is only one SN II-pec in the dataset, sn1987A \citep{kunkel_supernova_1987}. Thus we removed this SN and its 241 associated spectra.

We have also removed six other SN subtypes from the dataset: Ia-csm, Ib-pec, Ic-pec, IIL, IIn, and Ia-pec. The first five were removed because we had fewer than 50 spectra total after preprocessing. Ia-pec SN spectra were removed because we hypothesized that the `peculiar' SN subtypes were historically used as a catch-all `other' classification that was harming our model's performance.

After preprocessing, the number of spectra from each of the remaining ten SN subtypes is illustrated in \autoref{fig:treemap}. \autoref{table:data} contains the number of each spectrum present in the dataset at each stage of our data preparation. All data preparation stages are detailed in \autoref{sec:data:prep}.

\subsection{Pre-preprocessing} 
\label{sec:data:prep}
\label{sec:data:prep:prepreprocessing}

Any spectrum taken directly from a spectrograph is not yet suitable for training or classification. After processing the 2-D spectrum (\ie{} the image coming from the telescope) into a 1-D spectrum (a complex image processing task in and of itself), high-frequency noise in the spectrum is removed with a low-pass filter. A redshift estimate is obtained from the spectrum (\eg{} from \snid{}) and the spectrum is de-redshifted to its rest frame. The spectrum is re-binned to a consistent set of bins that are evenly spaced in log-space (\ie{} $\log{\lambda_i} - \log{\lambda_{i-1}}$ is constant).

Next, the continuum (the blackbody component to the spectrum) is removed by fitting and dividing by an N-point spline fit on the spectrum, where N is commonly around 10 to 13. \citet{blondin_determining_2007} notes that this removes all remaining color information from the spectrum. Continuum removal leads to artifacts at the edges of the spectrum, which can be smoothed by multiplying the edges by a cosine, a process known as apodizing. Finally, the spectrum is normalized (\eg{} to mean zero, or to between 0 and 1, or to however else is desired).

The preceding preprocessing is done by \snid{} and is detailed in \citet{blondin_determining_2007} {and also followed in many other related works including \dash{} itself,  \citealt{zhang2024maven}, etc.}

\subsection{Resampling spectra to low-resolution} \label{sec:data:prep:lowres}
The native resolution of each spectrum may vary, but when the spectra are re-binned by \snid{} they are all set to the common resolution of $R = 738$. This is considered high-resolution for SN classification, with bin-widths ranging from about 6\AA{} to 10\AA{} in the optical range. SNe are explosive processes and spectral features are broadened by the high velocity of the ejecta --- typically 1,000s of $\mathrm{km~s^{-1}}$. We will demonstrate that the performance of a classifier does not suffer when the spectra it is trained on are re-binned to a much lower resolution. In fact, the classifier may even perform better if re-binning suppresses noise while preserving signal. This will enable the application of memory-intensive neural network architectures like the attention mechanism, which is $O(N_{\rm wvl}^2)$ for the number of wavelength bins in the spectrum, $N_{wvl}$. \abcsn{} is trained on data that is lowered to $R = 100$.

\thesis{The first step in artificially lowering the spectral resolution from a high to low $R$ is to form the new set of wavelength bins which are defined by the new $R$; again, the new wavelength bins have a constant bin width in log-space.}

While we could simply re-bin our spectra to reduce memory costs, it is more interesting to answer the question ``What is the lowest resolution I need to collect my spectra at in order to responsibly use spectroscopic resources?'' After processing the spectra following the \snid{} steps, we convolve them with a Gaussian that serves as a generic approximation of a spectrograph's instrumental response in order to simulate a native low-resolution spectrum.

The Gaussian's Full Width Half Maximum (FWHM) is defined as a function of bin-width. That is, $$G(\mu_i, \sigma_i) = G(\lambda_i, \sigma(\Delta \lambda_i))$$ where $G$ is a Gaussian function with mean $\mu_i$ and standard deviation $\sigma_i$, $\lambda_i$ is the wavelength of the center of bin $i$, and $\Delta \lambda_i$ is the size of bin $i$. We choose ${\rm FWHM} = \Delta \lambda_i R_{\rm original} / R_{\rm lowered}$, and the standard deviation of a Gaussian is related to its FWHM by $\textrm{FWHM} = 2\sigma \sqrt{2 \ln{2}}$. We normalize $G$ such that its integral along the spectrum is unity so that the convolution does not add or remove flux from the spectrum.

\begin{figure}[t]
    \includegraphics[width=\columnwidth]{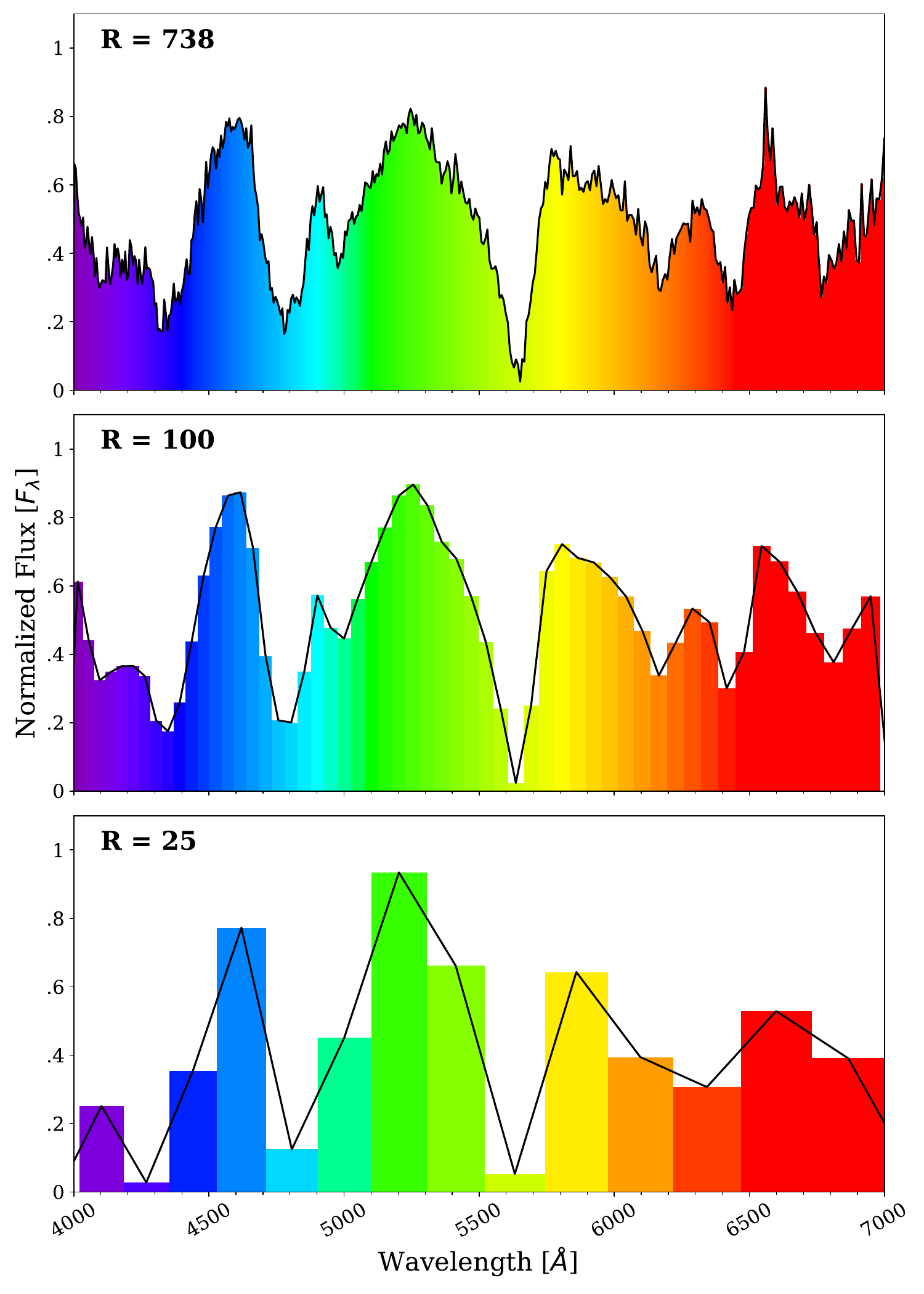}
    \centering
    \caption{Three spectra from SN1998dt, a type Ib SN, observed 1.8 days after peak brightness. The blackbody continuum has been removed. Type Ib SNe do not show hydrogen spectral lines nor the Si II$\lambda$6355 that characterizes SNe Ia. {\it Top:} The spectrum is plotted at the original high-resolution of $R = 738$. {\it Middle:} The spectrum is plotted at the low-resolution of $R = 100$, the same resolution that the \sedm{} operates at. {\it Bottom:} The spectrum is plotted at the ultra-low-resolution of $R = 25$.}
    \label{fig:example_specs}
\end{figure}

\begin{figure}[t]
    \includegraphics[width=\columnwidth]{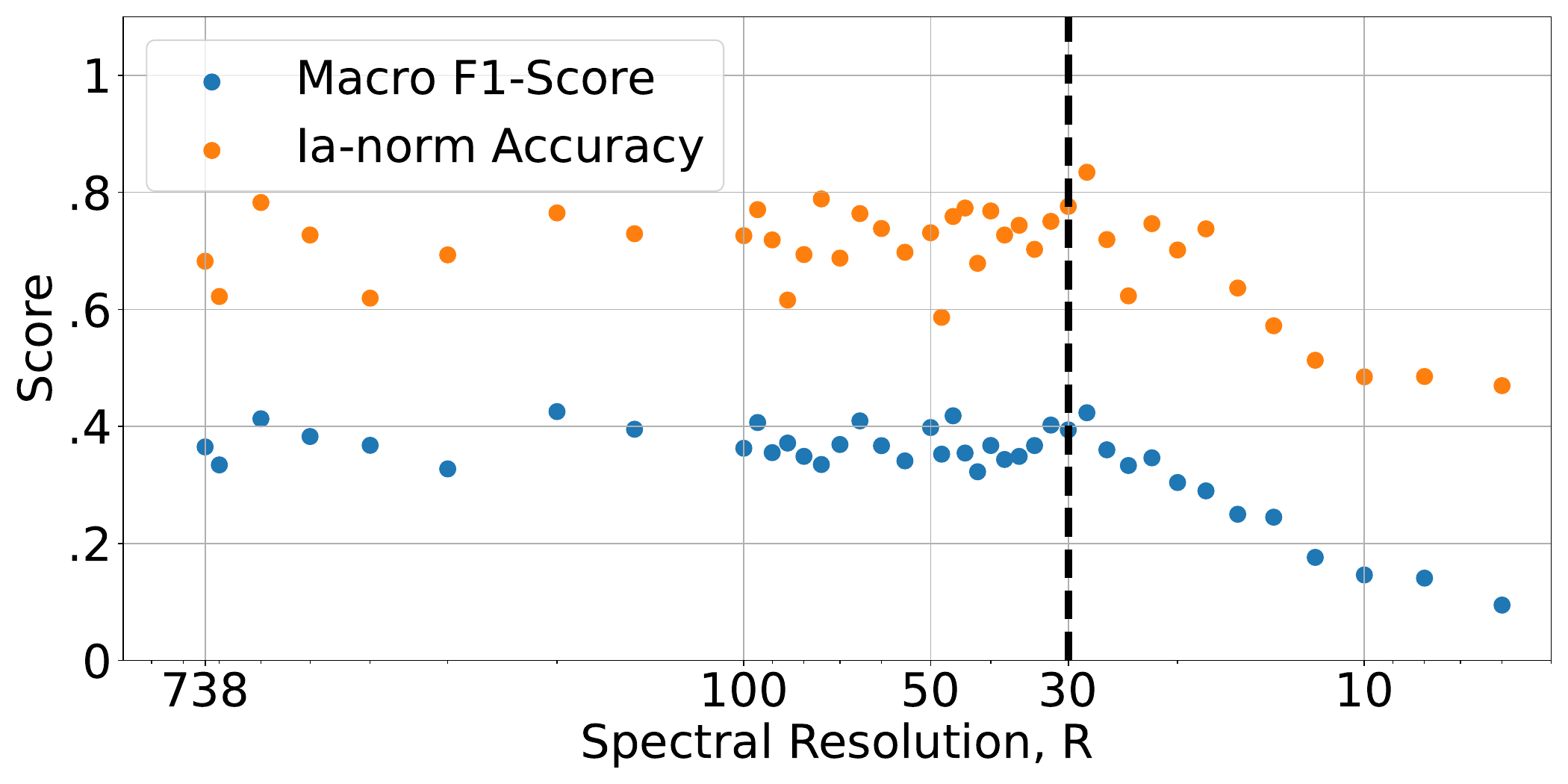}
    \centering
    \caption{This figure shows how the performance of DASH changes as a function of the spectral resolution of the dataset it is trained on. We lower the spectral resolution of our dataset to 40 different values between the native resolution, $R = 738$ and $R = 5$ in order to examine how the performance of DASH would change. In blue, the macro F1-score on the test set is plotted at 41 different spectral resolutions. In orange, the model's accuracy to predict SNe Ia-norm — the most numerous of the ten classes — is plotted. We qualitatively identify $R\approx{}30$ as the threshold below which DASH loses predictive power. Note that here \dash{} is trained with a decontaminated training-test split and early stopping to avoid overfitting (see \autoref{sec:retraindash}).}
    \label{fig:performance_vs_R}
\end{figure}

The last step in obtaining the low-resolution spectrum is to linearly interpolate the convolution (which is defined on the original spectrum's wavelength bins) to the low-resolution wavelength bins. \autoref{fig:example_specs} shows an example of a spectrum at the original $R = 738$, the resolution of the \sedm{} $R = 100$, and the ultra-low-resolution $R = 25$.

With this new low-resolution version of our dataset, we show that \dash{} does not suffer any loss in classification power down to $R \approx 30$. \autoref{fig:performance_vs_R} shows the performance of \dash{} at the resolutions between $R = 738$ and $R = 6$.

Further investigation of optimal resolution and signal-to-noise ratio for spectra collection will be the focus of a separate paper. In this work, we settle to perform our classification at a conservative $R = 100$ resolution.

\subsection{Culling, Standardizing and Validating} \label{sec:data:prep:culling_standardization}
Each spectrum is defined from 2,500\AA{} to 10,000\AA{}, but we set each spectrum's flux value to 0 at wavelengths shorter than 4,500\AA{} and longer than 7,000\AA{} to enforce uniformity. While this is a restricted wavelength range it served to ensure our data was maximally homogeneous and avoided working with missing values. We also normalize each spectrum to be mean $\mu=0$ and standard deviation $\sigma=1$, not including the 0-padding. At this point, we remove any spectra that have either a very small (0.1) or very large (100) dynamic range (the difference between a spectrum's minimum value and maximum value).

Our spectral dataset exhibits extreme class imbalance, with our most numerous class, Ia-norm, making up over $56\%$ of the dataset while the smallest class, Ibn represents less than $1\%$. Machine learning best practices dictate that it is important to stratify the train-test split —-- ensuring that if the training set represents $X\%$ of the total dataset, then $X\%$ of the occurrences of each class should be in the training set to avoid enforcing biases in the classifier (as discussed in \autoref{sec:retraindash}).

{On the other end, best practices would also entail a three-way data split with training-validation-test set with the latter unseen until final performance assessment or an N-fold cross-validation schema. However, it is simply not possible to split the data in three or more groups for us while retaining the small classes, because we only have three individual SN-Ibn, and six individual SN IIP and Iax. With the constraint of keeping all spectra from a SN in only one of the sets, splitting the data in more than two groups would make learning and assessment on small classes unrealistic.

We split our data into two approximately equal-size groups for training and testing,\footnote{As indicated in \autoref{sec:retraindash}, we follow \citep{bishop2006pattern}'s notation referring to the two subsets of our data as \emph{training} and \emph{test} set.} and} validate the stability of our results training \abcsn{} five times with different random seed initializations. The performance reported here is the mean performance of the five models. {We further confirmed the stability of the model by inverting our train and test set (essentially performing a 2-fold cross-validation), where we obtained a virtually identical performance. Finally, in \autoref{sec:results}, we will demonstrate that our model performance is preserved by predicting on an unseen dataset. Nonetheless,} due to the severe imbalance of our dataset (see \autoref{table:data}) the transferability of our result should be considered with caution for the smallest classes. There is no way to ensure that the classification power will remain as measured if the class is only populated by three (Ibn) or six (Iax, IIP) unique SNe as we simply do not have a sample large enough to assume our data is representative.

\thesis{In order to maintain confidence in the generalizability of the model, the training and test sets must not contain any information about each other whatsoever. When a model is trained on the same data that it is evaluated on, there is no reasonable basis that the model will generalize to new data. In other words, there is no way to determine if the model has `overfit' and learned the features of the particular dataset rather than the general features of the data. Overfitting typically leads the model to perform worse on the test set than the training set. However, if the sets are mixed, then this disparity will not arise, and overfitting will be very difficult to determine. If spectrum 1 from SN A was placed in the training set, and spectrum 2 from SN A was placed in the test set, this would constitute training set contamination where any machine learning model would be ``training on the answer''.}

\thesis{The training set would be what the model trains on. At each epoch of training, the loss function and other metrics would be calculated on the {\it test set} in order to monitor overfitting and generalizability during training. After the model is finished training, the results and metrics to be published would be calculated on the validation set.\footnote{The definitions of the test and validation sets are transposed by some. Some also call the three sets training, development and test.} Ideally, we would have this third split.}

\subsection{Data Augmentation} \label{sec:data:prep:augmentation}
\thesis{Class imbalance can be mitigated with data augmentation or oversampling. Data augmentation is the process by which new data is generated through various transformations of existing samples. Some common techniques for data augmentation are noise injection and affine transformations like translations, rotations, squeezing, stretching, and scaling.

Data augmentation does not increase the information content of a dataset, but using data augmentation to balance the classes is very important. Furthermore, data augmentation can help a model deal with noise (or any other natural phenomenon that resembles the augmentation techniques used), so it is often done even when class imbalance is not an issue. `Oversampling' is the simple cloning of samples in your dataset. This is sometimes done in order to balance the number of samples in each class exactly. `Oversampling' is also sometimes used synonymously with `data augmentation'.}

We employ three data augmentation techniques on our training set. First, we add some white noise with mean $\mu_\mathrm{noise}=0$ and a standard deviation $\sigma_\mathrm{noise}=0.10$ to each spectrum. Next, we randomly shift the spectra by at most five bins red or blue (up to 225\AA{} on the blue end and 350\AA{} on the red end), simulating inaccurate redshift corrections. Finally, we loosely simulate telluric lines by injecting 0 to 4 spikes along the (standardized) spectra. These spikes are distributed uniformly across the non-zero part of each spectrum. The magnitude of each spike is drawn from a $\mathcal{N}\sim(0,2\sigma)$ where $\sigma$ is the standard deviation of the spectrum itself. The sign of the spike --- \ie{} whether it is in emission or absorption --- has an 80\% chance to be in emission (telluric lines are of course, in absorption, but we commonly find artifact spikes in correspondence with the location of telluric lines caused by their removal). The details of these augmentation techniques are collated in \autoref{appendix:hyperparam}~\autoref{table:hyperparams}.

\begin{figure}[t]
    \includegraphics[width=0.80\columnwidth]{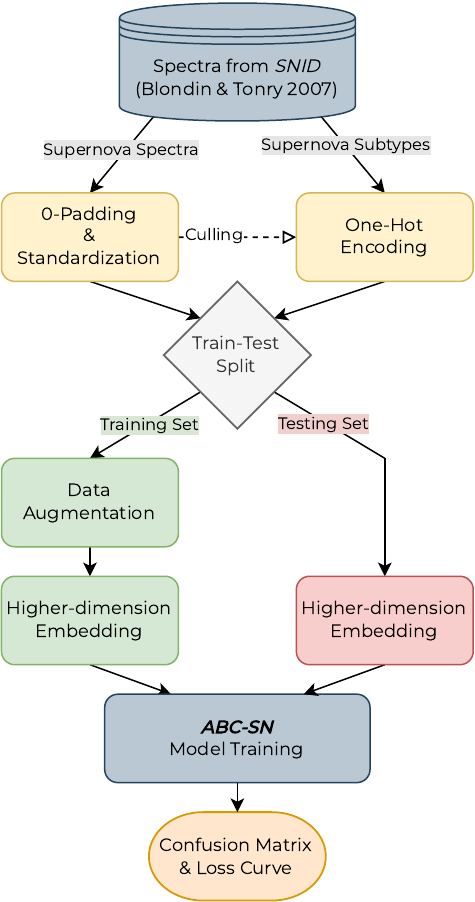}
    \centering
    \caption{A flowchart that summarizes the operations we perform on our spectral data. All preprocessing steps are described in \autoref{sec:data:prep}.}
    \label{fig:dataprepflowchart}
\end{figure}

In order to balance the dataset, we generate many copies of each spectrum and perform the augmentation described above separately on each copy. Enough copies of the spectra from each class are generated such that all classes have roughly equal size (see the fourth column in \autoref{table:data}). No oversampling or data augmentation is applied to the test set.
Our workflow is summarized graphically in \autoref{fig:dataprepflowchart}.

\section{Methods and Models} \label{sec:methods}
Here we describe the development of \abcsn{}, including all architectural elements and evaluation metrics.

\subsection{Model Performance Metrics}\label{sec:methods:metrics}
\thesis{A crucial aspect of machine learning is the choice of figure of merit for the model, often called the metric. Model accuracy is often the default choice for machine learning classification. However, accuracy is usually a misleading metric when working with imbalanced datasets.}
In the presence of an imbalanced dataset, appropriate performance metrics must be chosen to assess the performance of the model.

\thesis{Consider a dataset of 100 samples where 99 samples belong to class A and 1 sample belongs to class B. Now, consider a model that predicts class A no matter the input. This model will achieve 99\% accuracy without learning anything about the data. Additionally, if there were 100 new samples where 50 belonged to class A and B, then the model's accuracy would then be calculated as 50\%. Now consider two new metrics: `precision', defined as the number of true positives divided by the number of total positives predicted (the sum of true positives and false positives), and `recall', defined as the number of true positives divided by the number of samples that were predicted positive (the sum of the true positives and false negatives).}

Precision, also known as `positive predictive value', answers the question ``How many retrieved items are relevant?'' Recall, also known as `sensitivity', answers the question ``How many relevant items are retrieved?'' While accuracy can be calculated for all classes at once, precision and recall must be calculated for each class separately.

\thesis{Next, consider the `F1-score', which is the harmonic mean of precision and recall. For our toy model and dataset above, the precision and recall of class A is 99\% and 100\%, giving an F1-score of 99\%, while the precision and recall of class B are both 0\%, giving an F1-score of 0\% [Ignoring division by zero.].}

\begin{align*}
    \textrm{Precision} &= \frac{TP}{TP + FP} \\
    \textrm{Recall} &= \frac{TP}{TP + FN} \\
    F_1 &= \frac{2TP}{2TP + FP + FN}
\end{align*}

By calculating the F1-score for each class in a dataset and taking their {\it unweighted} average, we arrive at the `macro F1-score'. The macro F1-score provides a much more balanced picture of the model performance than, \eg{} accuracy, when dealing with an unbalanced dataset. It is for this reason that we consider the macro F1-score to be the most appropriate metric when comparing models for this analysis. It is the F1-score of the test set that we monitor during training.

\thesis{The macro F1-score for the toy model and dataset is $\sim 50\%$, equivalent to flipping a coin for this binary classification.}

\subsection{ABC-SN: An Attention-Based Classifier}\label{sec:methods:abcsn}
The architecture of \abcsn{} is broadly inspired by the Transformer \citep{vaswani_attention_2017}. Modifications are needed to apply this model, natively built for Natural Language Processing (NLP), to spectral data. Here we describe how we adapted each element of the architecture inherited from the Transformer, including a discussion of choices that we ultimately abandoned, so as to aid researchers engaging in similar projects. \abcsn{} is implemented in TensorFlow Keras \citep{chollet2015keras}.

\thesis{The Transformer \citep{vaswani_attention_2017} was not designed for the inferential machine learning task of classification, so it must be adapted. The heart of the model is the `encoder block' and `decoder block'. The former encodes and learns information about the inputs, the latter decodes this information and generates output. The decoder is for generative machine learning, \eg{} translating sentences or answering questions, and is unnecessary for inferential tasks.}

We must clarify immediately that we only include the encoder elements of the Transformer in our architecture since our purpose is inferential, rather than generative.

\begin{figure}[t]
    \includegraphics[width=\columnwidth]{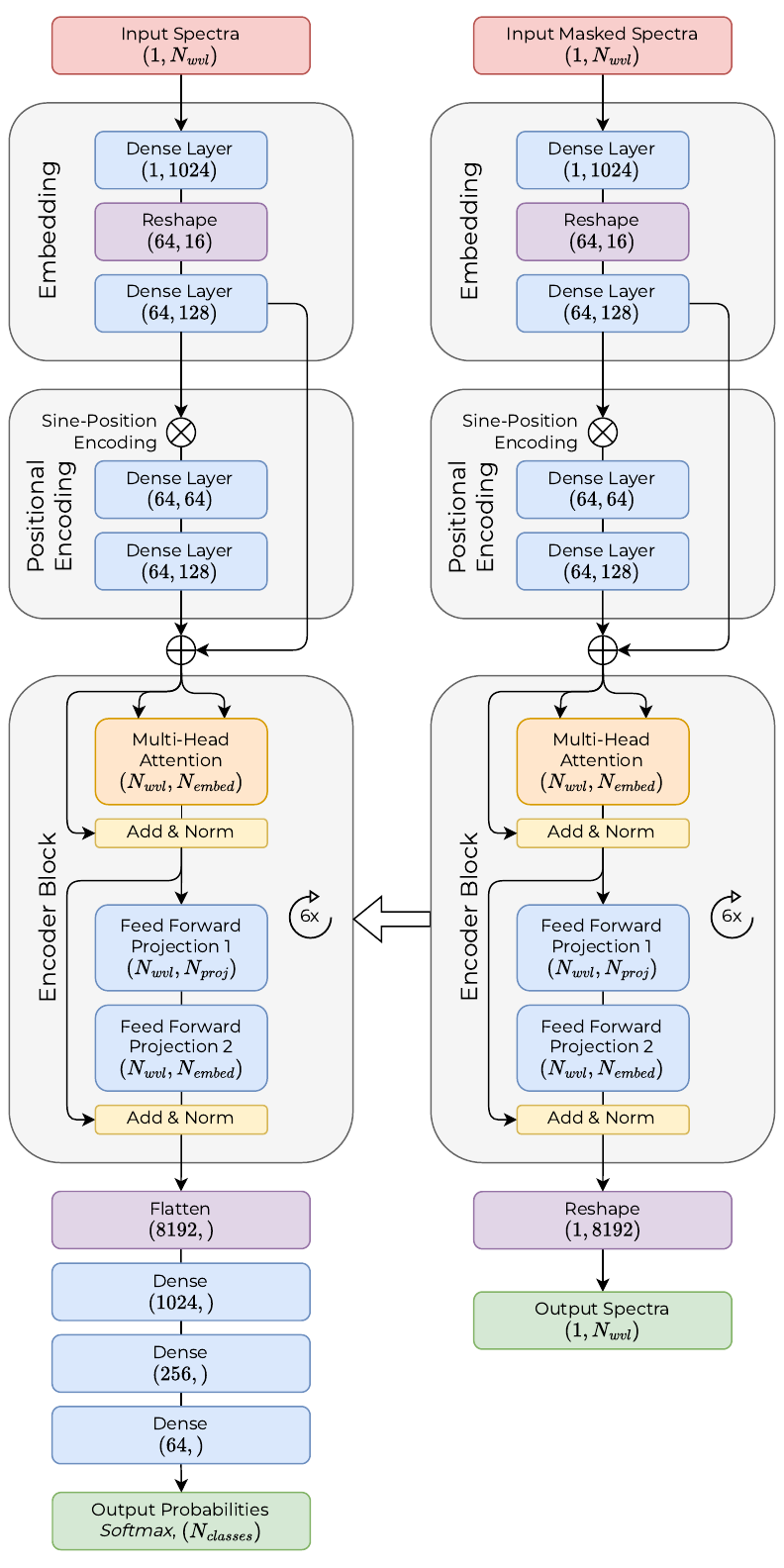}
    \centering
    \caption{The conceptual diagram of the \abcsn{} architecture described in \autoref{sec:methods:abcsn}.}
    \label{fig:abc_architecture}
\end{figure}

The architecture of \abcsn{}, illustrated in \autoref{fig:abc_architecture}, begins with the input spectra, assumed to have shape $(1, N_{\rm wvl})$ where $N_{\rm wvl}$ is the number of wavelength bins that a spectrum is defined on. These spectra are then embedded into a higher dimension, described in \autoref{sec:methods:abcsn:embedding}. Next, positional information is encoded into the spectra (\autoref{sec:methods:abcsn:positionalencoding}) before being passed to the encoder block (\autoref{sec:methods:abcsn:encoder}). Finally, data passes to the feed-forward classification head (\autoref{sec:methods:abcsn:ffhead}), which provides the final output classification probabilities.

\begin{figure*}[h!]
    
    \includegraphics[width=0.7\textwidth]{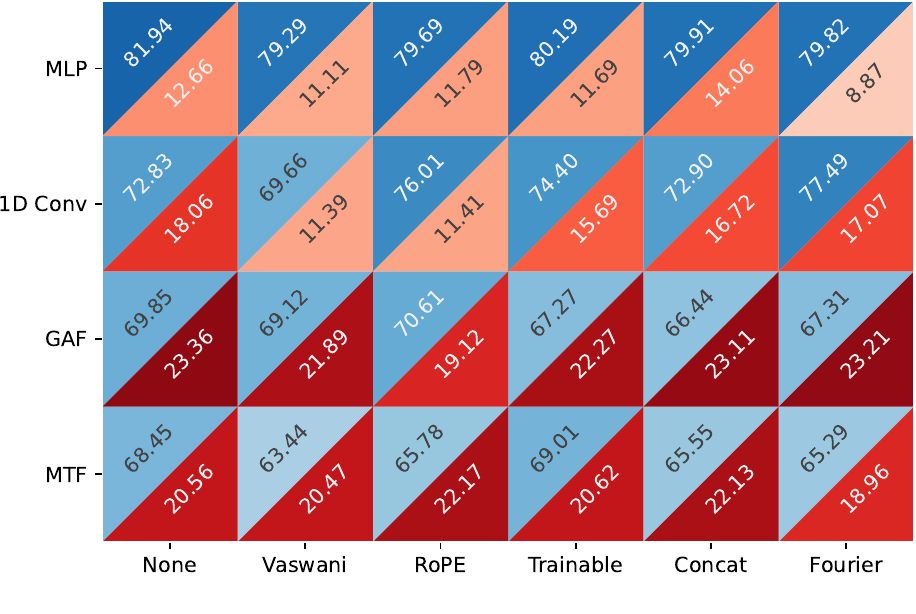}
    \includegraphics[width=0.7\textwidth] {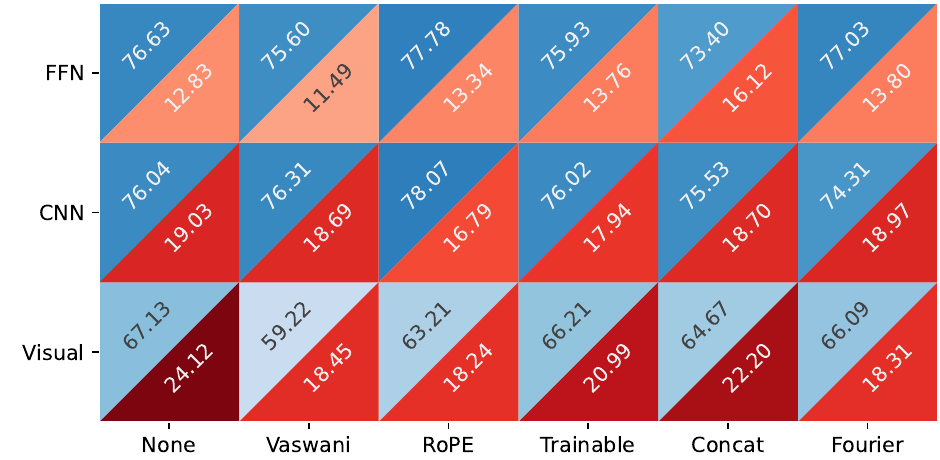}
    \centering
    \caption{Results of tests performed on different architecture element combinations. {\it Top}: each box corresponds to a combination embedding ($y$ axis) --- positional encoding ($x$ axis) as labeled and described in sections~\ref{sec:methods:abcsn:embedding} and ~\ref{sec:methods:abcsn:positionalencoding}, respectively and average over the three tested options for classification heads (see \autoref{sec:methods:abcsn:ffhead}).  {\it Bottom}: each box corresponds to a combination classification head ($y$ axis) --- positional encoding ($x$ axis) as labeled, average over the four embeddings. 
    The top triangle in each cell corresponds to the test set F1-score, the bottom to the amount of overfitting, defined as the difference in F1-score between training and test sets. The selected architecture for \abcsn{} corresponds to the top right cell of these plots: MLP embedding and Fourier positional encoding, Feed Forward (FF) classification head. While this combination does not achieve the highest F1-score it is at most two percentage points off from the best performer and shows the least amount of overfitting (colors: darker blues indicate higher F1-score, darker reds indicate more overfitting).}
    \label{fig:heatmap}
\end{figure*}

\subsubsection{The Embedding}\label{sec:methods:abcsn:embedding}
From the observatory to our database, a spectrum is preprocessed into a 1-D vector (see \autoref{sec:data:prep:prepreprocessing}). However, the attention mechanism was not designed with 1-D inputs in mind --- the positional encoding is typically applied by concatenation, sum, or dot-product, so the first action in our model is to increase the dimensionality of a spectrum vector.

We tested five different embeddings to take our 1-dim spectra into a higher-dimensional space that would be more friendly to the Transformer architecture. The first was to bin each flux value in the spectra according to what quantile that flux value was in within the distribution of that spectrum. For example, suppose we decide to embed into $N_{\rm embed}$ dimensions. We would histogram the flux values of the spectrum into $N_{\rm embed}$ bins, and we would populate a $(N_{\rm wvl}, N_{\rm embed})$ array with ones according to which bin each flux value falls into. This method proved not significantly better than having no embedding at all.

We then applied two methods from {\tt pyts}, a \texttt{Python} package for time series classification \citep{JMLR:v21:19-763}. The Gramian Angular Field (GAF) and Markov Transition Field (MTF) are images obtained from time series and we used a {\tt pyts} implementation of a function to calculate a GAF and MAF from our 1-D spectra. We transformed each spectrum into a GAF or MAF before passing it to \abcsn{} for training. However, we found that both the GAF and MAF generally performed worse than having no embedding.

The fourth embedding method is a learned embedding which uses a {\tt Conv1D} layer with a kernel size of 1 and $N_{\rm embed}$ filters to project the 1-D spectra into an $N_{\rm embed}$-dimensional space.

The fifth embedding method, the one that we adopted for \abcsn{}, is also a learned embedding. Inspired by {\it Astroconformer} \citep{pan_astroconformer_2024}, we use a multi-layer perceptron (MLP) to embed the spectra into a higher dimension. The first layer of the MLP is a {\tt Dense} layer with 1024 neurons and a linear activation function. This reshapes the spectra from  $(1, N_{wvl})$ to $(1, 1024)$. Next, we reshape the tensor to $(64, 16)$ and pass this to another {\tt Dense} layer with 128 neurons and a {\tt ReLU} activation function, leading to a final data shape of $(64, 128)$. This tensor is passed to the positional encoding. This embedding is illustrated in \autoref{fig:abc_architecture}.

\subsubsection{The Positional Encoding}\label{sec:methods:abcsn:positionalencoding}
\thesis{The attention layer is ultimately a simple dot product, which is an operation that takes two vectors and returns a scalar. When attention is used for NLP tasks, the input vectors are (tokenized) sentences of some Earthly language which obviously have a well-defined and syntactically important order; \ie{} the beginning of the sentence comes before the end of the sentence. But when the attention mechanism is applied to these vectors, this ordering is lost as the vector inputs are reduced to a scalar. SN spectra have a strictly defined ordering given by the wavelengths that correspond to each flux measurement. Additionally, the ordering of the spectral features is relevant to classification. In order to retain this information through the model network, we use a positional encoding (PE).}

A key element of the Transformer architecture, and of \abcsn{}, is the positional encoding (PE) which conveys the wavelength information for a given flux to the attention layers. A PE can be fixed or have learnable weights \citep{gehring_convolutional_2017, ahmed_transformers_2022}. We looked to \citet{moreno-cartagena_positional_2023} and \citet{pan_astroconformer_2024} for their work on the attention mechanism applied to photometric light curves in order to find a PE to implement in \abcsn{}. We tested six different PEs: No PE; ``Vaswani'' the PE used in \citet{vaswani_attention_2017}; ``RoPE'' or Rotary PE, used in \citet{pan_astroconformer_2024}; ``Fourier'', ``Trainable'', ``Concat'' all described in \citet{moreno-cartagena_positional_2023}. Based on a comparative test between all six methods, we decided on the ``Fourier'' PE for \abcsn{} as it offers the best balance of performance and overfitting. The results of the comparative architecture test can be seen in \autoref{fig:heatmap}, marginalized over three classification heads (described in \autoref{sec:methods:abcsn:ffhead}).

The Fourier PE is based on the PE used in \citet{vaswani_attention_2017} but modulates that PE with a hidden {\tt Dense} layer with a {\tt GeLU} activation function. The output of this hidden layer is then supplied to another {\tt Dense} layer with a linear activation function. We chose the hidden layer to have 64 neurons, while the number of neurons in the output layer is chosen such that the output shape matches the input shape. The PE is then added to the spectral data just before being passed to the first {\tt MultiHeadAttention} layer. This PE is illustrated in \autoref{fig:abc_architecture}.

\subsubsection{The Encoder Block}\label{sec:methods:abcsn:encoder}
The encoder block is composed of two sub-blocks: the attention sub-block and the projection sub-block, explained graphically in \autoref{fig:abc_architecture}. The first sub-block includes the eponymous {\tt MultiHeadAttention} (MHA) layer, which accepts two arguments to calculate attention on. Since we only use the encoder from the Transformer architecture, we are interested in self-attention, as per \citet{vaswani_attention_2017}, so we supply the layer with two copies of the inputs. The inputs to the MHA layer are directly added to its outputs, and this sum is normalized with a {\tt LayerNormalization} layer. The summing of inputs and outputs out of sequence is called a `residual connection`. Note that this sub-block does not change the shape of the data passing through it.

The second sub-block projects the data from shape $(N_{wvl}, N_{embed})$ to $(N_{wvl}, N_{proj})$. It should be the case that $N_{proj} > N_{embed}$ based on the original reasoning for the projection layer outlined in Section~3.3 of \citet{vaswani_attention_2017}, though it is technically possible for it to be any positive integer. We implement this projection sub-block with a {\tt Conv1D} layer with a kernel size of 1, $N_{proj}$ filters and a {\tt ReLU} activation function. The data is then projected back down from $(N_{wvl}, N_{proj})$ to $(N_{wvl}, N_{embed})$ with another {\tt Conv1D} layer with a kernel size of 1, $N_{embed}$ filters and a {\tt ReLU} activation function. Just like the attention sub-block, we use a residual connection between the inputs and outputs of the projection sub-block and then normalize it with a {\tt LayerNormalization} layer. Once again, this sub-block does not change the shape of the data passing through it.

The encoder block can be repeated as many times as desired, passing the outputs of one encoder block as the inputs of the next. Because the shape of the data going into the encoder block is preserved, there is no hard limit to the number of encoder blocks in a model. Providing each encoder block after the first with a residual connection of the PE by itself may help the model retain information about relationships between individual flux values. Encoder blocks may also be given a residual connection to the original spectral data or even to the output of one or more preceding encoder blocks.

We use a {\tt Keras-NLP} implementation of the transformer encoder block, and we use a dropout value of 50\%.

At this point in the architecture of the model, we hope that it has learned important information about how flux values in a spectrum relate to one another. Generally speaking, this is the information that a professional SN scientist would make a classification on and so the following neural network is designed for the same task.

\subsubsection{The Feed-forward Classification Head}\label{sec:methods:abcsn:ffhead}
The last element of the architecture of \abcsn{} is the classification head. The data exiting the encoder block will be 2-D. To transform it to the final 10-neuron 1-D layer representing probabilities for each class, we use three {\tt Dense} layers with $N_{FF_1}$, $N_{FF_2}$, and $N_{FF_3}$ neurons respectively, each with a {\tt ReLU} activation function. Finally, the model concludes with a {\tt Dense} layer with a Softmax activation function and ten neurons, one for each SN subtype in the dataset.

Each {\tt Dense} layer is given an {\tt L2} regularization of 0.01, and each {\tt Dense} layer is followed by a {\tt Dropout} layer with a dropout of 50\%.

\thesis{We choose to immediately perform this dimensionality reduction with a {\tt Reshape} layer.  Using a combination of 2-D convolutional and 2-D pooling layers would also be an effective, but more advanced, way to reduce the dimensionality. We tried this and found it did not improve the performance of the model. }

\thesis{When designing a network with {\tt Dense} layers, it's easy to include too many layers with too many neurons. A network like this would easily overfit and perform superbly on the training set but would have little generalizability and score poorly on the test set.}

\thesis{Applying a Softmax activation function ensures that all ten neurons in the layer are normalized to sum to unity, meaning that we can interpret the output of this layer as probabilities for classification.}

\thesis{It is common to declare a threshold for probabilistic classification, say 70\%, and if a model predicts a class with a probability above this threshold, then that is the class that is assigned to the input spectrum. This also enables one to define a `None of the above' class for any input spectra where the model does not predict any class above the threshold. One would also want to use a Sigmoid activation function in this case. . For simplicity, we did not use thresholding, but instead we take the highest probability class as the predicted class.}

Two alternatives for this classification head were trialed during \abcsn{} development. The first was a CNN hybrid model. The output of the encoder was passed to several convolutional layers before being passed to the three-layer feed-forward network previously described. This constituted a significant increase in model complexity with an insignificant change in performance compared to the feed-forward only classification head.

The second alternative is an adaptation of the visual attention model presented in \citet{yan_melanoma_2019}, which was designed for automated melanoma recognition. Their model uses \vgg{} \citep{simonyan_very_2015}, a widely used convolutional network, as a backbone and then incorporates attention layers. For our trials, this visual attention model also replaces the encoder portion of \abcsn{} and is overall more complex, though still relying on {\tt Conv2D} and {\tt MultiHeadAttention} layers. This architecture did not improve the performance.

The results of the comparative trials between these basic model architectures is reported in \autoref{fig:heatmap}.

\begin{figure}[t]
    \includegraphics[width=\columnwidth]{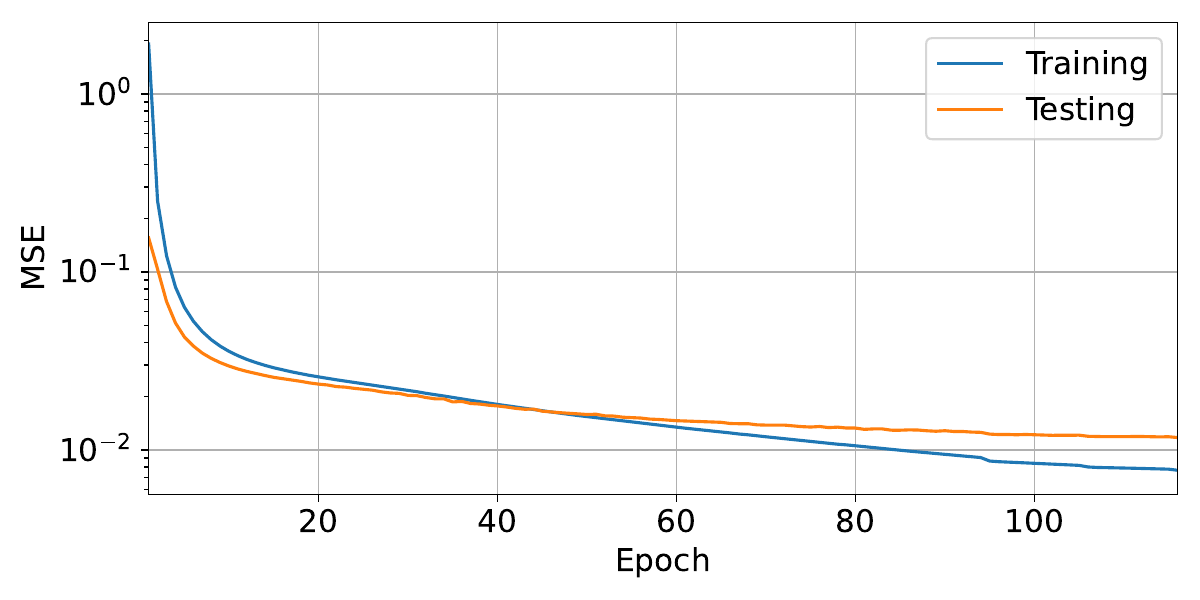}
    \centering
    \caption{The loss curve during \abcsn{} pretraining. The mean squared error (MSE) is shown for the training and test sets during pretraining. The model was pretrained for 141 epochs on reconstruction of masked spectra (see \autoref{sec:methods:pretraining}) before reverting back to epoch 116 due to an early stopping callback.}
    \label{fig:pretrainingloss}
\end{figure}

\subsection{Pretraining Attention Weights on Masked Spectra}\label{sec:methods:pretraining}
\thesis{It was imperative that we design \abcsn{} in a way that minimizes overfitting. With an unbalanced dataset of ten different classes, overfitting is likely to be an issue.}

One of the major steps we took to combat overfitting was to train the model in two stages. Inspired by masked language models \citep{devlin_bert_2019}, the first stage `pretrains' the weights of the encoder block on the regression task of predicting masked spectral values, while the second stage trains \abcsn{} with the weights in the encoder layers initialized from the pretraining.

We do this by constructing an \abcsn{} model that is truncated after the encoder blocks, shown on the right in \autoref{fig:abc_architecture}. The output of the encoder blocks is flattened and then connected to the output layer, which is a {\tt Dense} layer with a linear activation function and the layer bias turned off. The inputs {\it and} targets for this pretraining model are SN spectra, so the number of neurons in the final {\tt Dense} layer is equal to the size of the inputs.

The target spectra for the pretraining model are untouched, but the input spectra are masked and perturbed. A random, continuous 15\% of the non-zero portion of each spectrum is selected and set to 0. An additional 2.5\% of points on the non-zero portion of each spectrum is randomly perturbed by multiplying it with a Gaussian random variable with mean 0 and standard deviation 1. The weights of this model are trained to predict masked spectra values in order to generate a rich internal representation of the data. 

The loss function for this pretraining stage is shown in \autoref{fig:pretrainingloss}. At the end of pretraining, the weights for all of the encoder sub-layers are used to initialize the weights for those layers in the full \abcsn{} model.

\begin{figure}[t!]
    \includegraphics[width=\columnwidth]{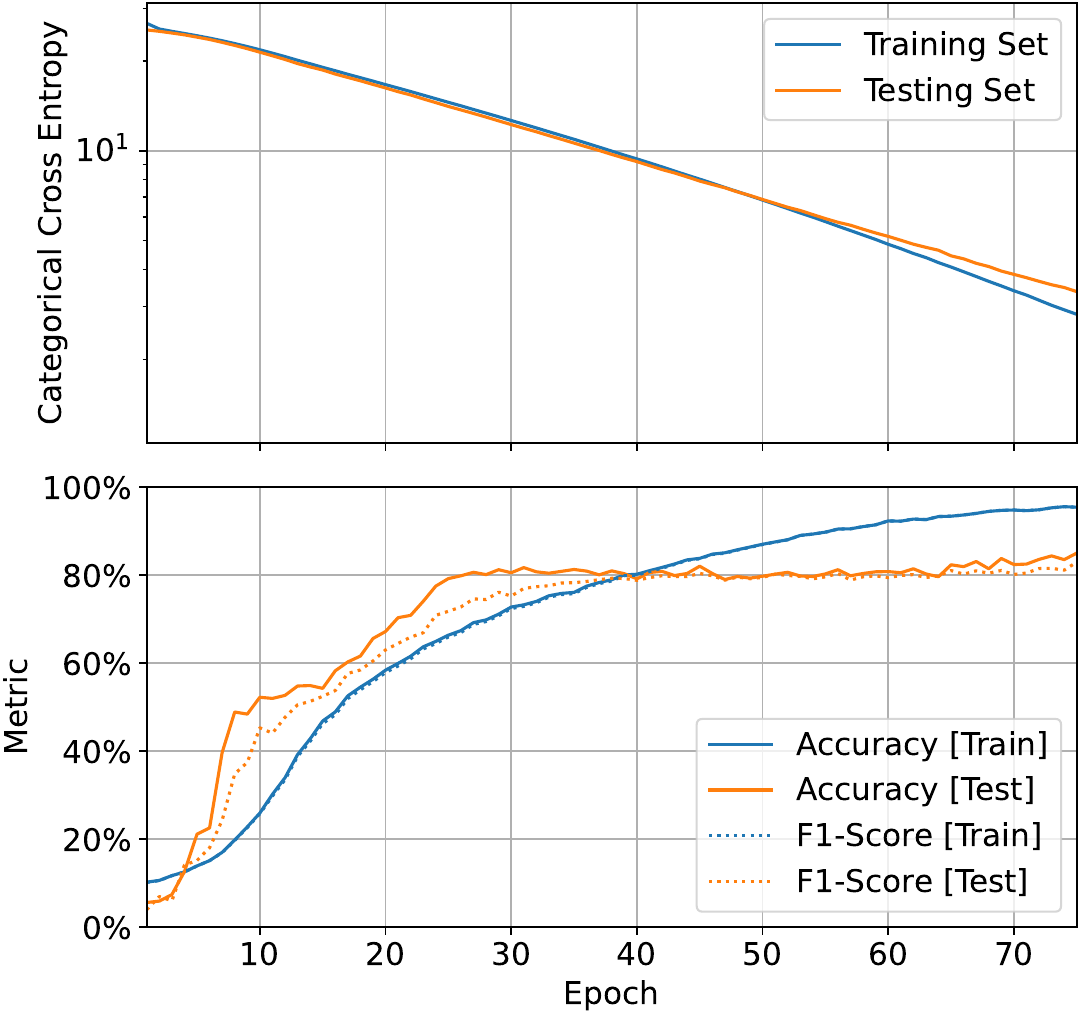}
    \centering
    \caption{The loss and metric curves for \abcsn{} during training. \textit{Top:} The categorical cross-entropy loss. \textit{Bottom:} The categorical accuracy (solid lines) and macro F1-score (dotted lines). \abcsn{} was set to train for 100 epochs before reverting back to epoch 75 due to an early stopping callback. The rapid rise of the training set metrics beyond the test set demonstrates this data's eagerness to overfit despite our significant efforts in regularization to prevent this. }
    \label{fig:abcsnloss}
\end{figure}

\begin{figure*}[h!]
    \includegraphics[width=\textwidth]{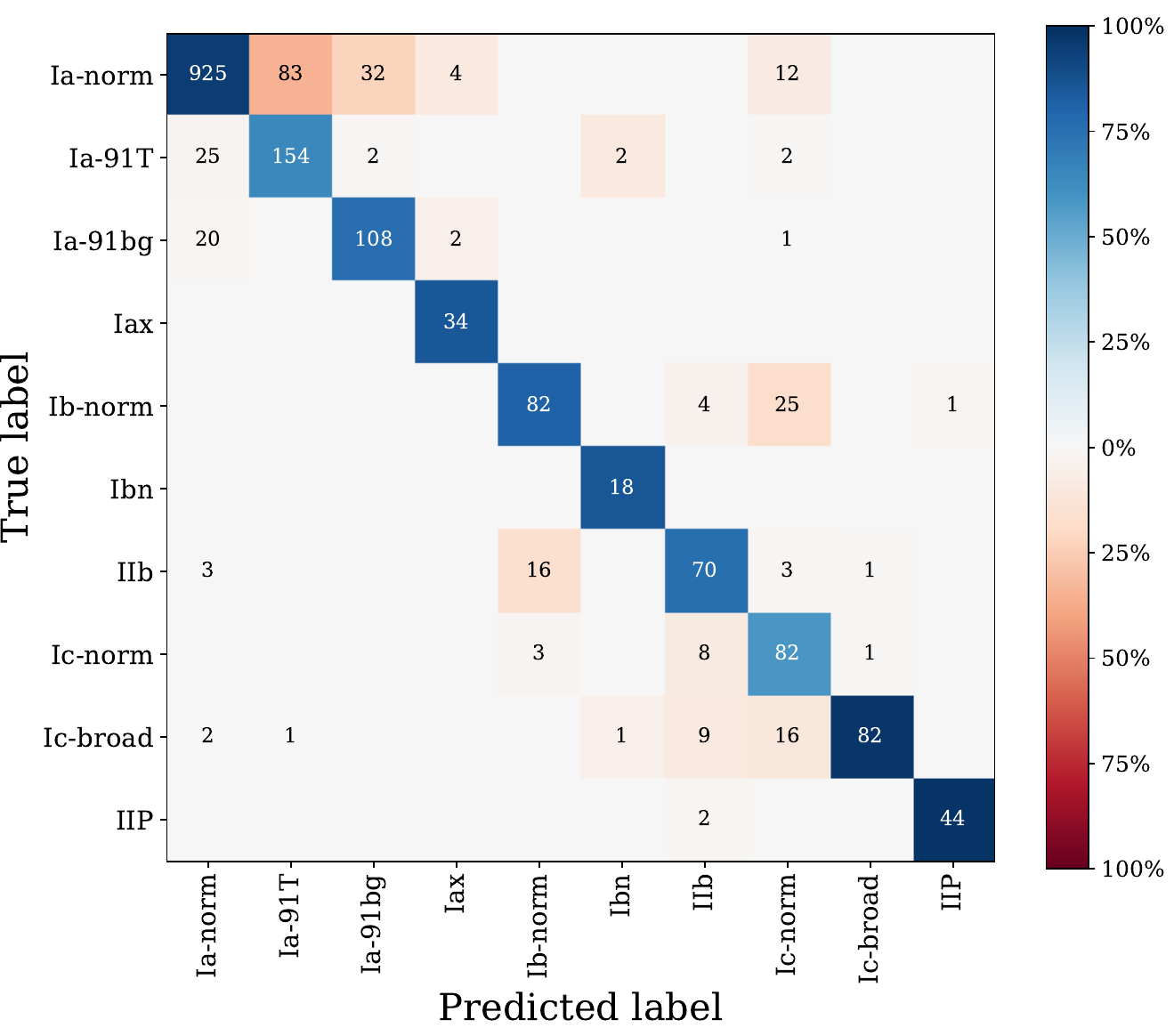}
    \centering
    \caption{Confusion matrix showing the performance of \abcsn{} in the classification of SNe in the SNID dataset. On the horizontal axis, the predicted class. On the vertical axis, the label. As in figure \autoref{fig:dash_compare}, cells on the diagonal (correct classifications) are plotted in shades of blue and off-diagonal (incorrect) in shades of red corresponding to the percentage (in)accuracy. The percentage is calculated along the columns (emphasizing completeness). In each cell, the number of objects with the corresponding prediction and label is reported, when that number is larger than 0 (note the significance class imbalance in the test set, discussed in \autoref{sec:data}). This confusion matrix shows the clear and significant improvement over the state of the art, \dash{} (see \autoref{fig:dash_compare}, right top panel) on the classification of 10 selected subtypes. 
    The model performance is discussed in detail in \autoref{sec:results}. A set of confusion matrices normalized by column (emphasizing completeness) and by row (emphasizing purity) are available in \autoref{appendix:cm} with percentage performance reported in each cell.}
     \label{fig:THE_cm}
\end{figure*}

\subsection{Computation Costs}
\abcsn{} was trained on one Nvidia T4 GPU for 15 minutes. The pretraining model was trained on the same GPU for 10 minutes. With the same hardware, \abcsn{} takes $242\pm 4.19\ {\rm ms}$ to predict on the test set (1880 spectra, $128\ \mu s$ per spectrum).

We retrained \dash{} with 16 CPUs and 64GB RAM. When retraining \dash{} with the original training schema (see \autoref{fig:dash_compare}, left), training lasted $~36$ hours (576 CPU hours). When the early stopping protocols are enabled, \dash{} training lasted 45 minutes (12 CPU hours; see \autoref{fig:dash_compare}, right). With the same hardware, \dash{} retrained to avoid overfitting takes $8.07\ {\rm s}\pm 79.8\ {\rm ms}$ to predict on the test set ($143,920$ spectra, $56\ \mu s$ per spectrum).

\section{Results and Discussion} \label{sec:results}
\thesis{We have identified four aspects of \abcsn{} performance that are most important for determining its overall usefulness. First, we must examine to what degree \abcsn{} overfits in order to confidently proclaim that the model's performance will be applicable to new data. Next, we can calculate metrics like accuracy, F1-score, purity, and completeness, which will help astronomers determine whether \abcsn{} is useful for their specific science case. Finally, we can make comparisons to \dash{} as the current state-of-the-art SN classifier.}

To ensure \abcsn{}'s performance would extend to newly discovered SNe we evaluate overfitting by monitoring the model loss at every training epoch on both training and test sets. \autoref{fig:pretrainingloss} shows the mean squared error (MSE) loss during \abcsn{} pretraining (masked model predictions, see \autoref{sec:methods:pretraining}). \autoref{fig:abcsnloss} shows the categorical cross-entropy loss during \abcsn{} training (classifier model). During both training and pretraining, we use an early stopping callback to prevent overfitting and avoid wasting computation resources (see details in Appendix, \autoref{tab:hyperparams}). \abcsn{} trained for a total of 100 epochs when the early stopping callback was activated; the final model represents 75 trained epochs.

The final accuracies on the training and test sets are $92.12\%$ and $83.81\%$. The final macro F1-scores are $92.08\%$ and $82.45\%$. \dash{} accuracies on the training and test sets are $67.84\%$ and $61.97\%$; \dash{} macro F1-scores are $77.06\%$ and $58.86\%$.

\begin{table}[t!]
    \centering
    \begin{center}
    \begin{tabular}{l|llll}
    \small
    & \multicolumn{2}{c}{Completeness} & \multicolumn{2}{c}{Purity} \\
    SN Subtype & \abcsn{} & \dash{} & \abcsn{} & \dash{}  \\
    \hline    
    {\bf Ia}&{\bf86.8\%}&	{\bf63.2\%}&	{\bf88.9\%}&	{\bf83.0\%}\\
    Ia-norm & 87.6\% & 61.6\% & 94.9\% & 92.5\% \\
    Ia-91T & 83.2\% & 76.2\% & 64.7\%  & 23.5\% \\
    Ia-91bg & 82.4\% & 50.4\% & 76.1\%  & 66.0\% \\
    Iax &  100\% & 94.1\% & 85.0\% & 58.5\% \\
    \hline
    {\bf Ibc }&	{\bf78.0\%}&	{\bf53.1\%}&	{\bf79.3\%}&	{\bf52.4\%}\\
    Ib-norm & 73.2\% & 60.5\% & 81.2\% & 61.9\% \\
    Ibn & 100\% & 25.0\% & 85.7\% & 100\% \\
    IIb & 75.3\% & 44.8\% & 75.3\% & 59.1\% \\
    Ic-norm & 87.2\% & 59.1\% & 58.2\% & 25.7\% \\
    Ic-broad & 73.9\% & 51.9\% & 97.6\% & 51.9\%\\ 
    \hline
    IIP & 95.7\% & 95.3\% & 97.8\% & 67.2\% 
    \end{tabular}
    \end{center}
    \caption{Per subtype accuracy of \abcsn{} compared to \dash{}, retrained on this dataset with updated training-test split and preventing overfitting as discussed in \autoref{sec:retraindash}. The first two columns show per-class prediction completeness, the last two purity. In all cases, the predicted class is assigned by selecting the class with the highest prediction probability (but see \autoref{sec:results} and  \autoref{fig:roc} for results based on probability thresholds).}   \label{tab:results}
\end{table}

A confusion matrix showing per-class performance is shown in \autoref{fig:THE_cm} (and additional confusion matrices showing precision/purity and recall/completeness are included in \autoref{appendix:cm}). \autoref{tab:results} shows the performance for each subtype for both \abcsn{} and the \dash{} benchmark model.

For nearly all classes, \abcsn{} outperforms \dash{}. Notably, it outperforms \dash{} by 26\% points on SN Ia completeness and 2.4\% in SN Ia purity. The only exception is purity on the SN Ibn sample, but the small size of the sample (27 viable spectra of 3 SNe, only 1\% of the test set --- see \autoref{table:data} and \autoref{fig:treemap}) makes estimates of the performance on this subclass unreliable. {Aggregating over broad classes: SNe Ia, comprehensive of all subtypes, improve by $>23\%$ points in completeness and nearly 9\% in purity; for SN Ibc (all stripped envelope SN subtypes) completeness increases from 53.1\% with \dash{} to 78\%, with \abcsn{}, $\sim25\%$ points, and purity by $\sim27\%$ points}.

\begin{figure}
    \centering
    \includegraphics[width=\linewidth]{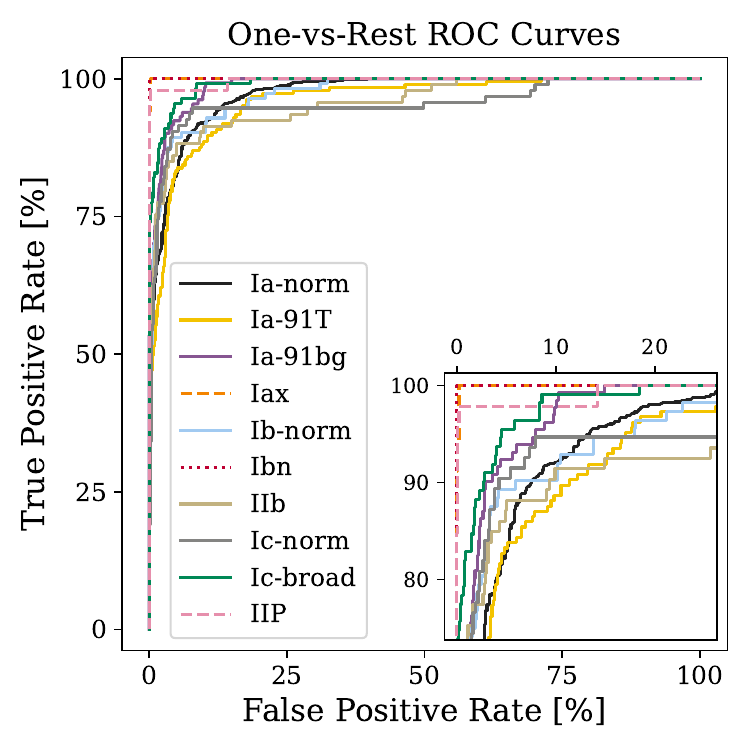}
    \caption{{ROC curve for ABC-SN One-vs-rest classification for each considered class. The insert shows a zoom in of the top left region of the ROC space. We note that the smallest classes (IIP and Ib-n, 6 objects, and Iax, 3 objects, dashed or dotted lines) stand out in the ROC curve as the extremely small dataset prevents resolution in the result.}}
    \label{fig:roc}. 
\end{figure}

{Our classification results are based on selecting the highest predicted class as our final prediction, with no cuts on probability. If the user desires to enhance class purity, for example, to automate classification without human vetting, or completeness, they could set a probability threshold such that a prediction is only made if the threshold is passed. To that end, we provide ROC curves for each class as a one-vs-rest classification result (\autoref{fig:roc}). For example, at classification threshold $p=0.92$ we obtain $>3\%$ purity on SN Ia with a completeness of $>76\%$ while for all other classes, purity is better than $3\%$ and completeness better than $65\%$ (IIb and Ia-91T have the lowest completeness). At a threshold that gives $>97\%$ completeness for SN Ia, purity is $17\%$.}

{The reader also may wonder how the results depends on spectral epoch since our direct comparison model was designed to predict spectral epoch jointly with type and SN spectral characteristics indeed evolve over time \citep[e.g.,][and many more]{yesmin2025spectral, 2023A&A...677A...7D, gutierrez2017type, drell2000type} in some cases so significantly to transition between root taxonomical classes (\eg{} SN IIb). Nonetheless, we find our results are stable across time intervals. Our training data includes spectra between -20 and 50 days of maximum light (\autoref{sec:data}). We stratified our test data after training into three bins: {\it pre-peak}, -20 to 10 days of maximum light, {\it near-peak} -10 to +20 days, and {\it post-peak}, 20 to 50 days.
For the near-peak spectra (1,311 spectra), compared to the full testing set (1,880 spectra), our precision, recall weighted by subtype remains constant, while the accuracy increases by one percentage point, as does the F1 average score. For the post-peak sample (496 spectra) scores all increase by one percentage point. Unfortunately, only 73 spectra are available in the range -20 to -10 subset, so a robust result cannot be reported; we note that we see a drop in performance, but this is largely dominated by multiple subclasses being entirely empty (confusion matrices for the near-peak and post-peak samples can be found in  \autoref{appendix:cm}.}

Our final performance test was done on data unseen during training using the data in \cite{magill2025super} and 14 selected spectra of normal, nearby SN Ia's. While this test was performed to ensure the transferability of our model, we reserve more extensive testing on larger and more diverse samples to future work. First, we useddata already pre-processed according to the standard SNID steps \autoref{sec:data:prep}, in the -10 to +20 days to peak range used in the phase stratification tests just discussed (see also \autoref{appendix:cm}). We include spectra from all new in \cite{magill2025super} SNID template objects that were not included in our original training and test data (see \autoref{appendix:qub}) and classify them with the pretrained model weights\footnote{\url{10.5281/zenodo.16620816}}, leading to the results shown in \autoref{tab:superfitperformance}.  While the small and imbalanced sample induces, once again, some noise, this test confirms our performance and the transferability of our model. A score for SN Ia-norm and Ib-norm is not reported. In \autoref{appendix:qub} we report the label and predicted classification for each object in the unseen sample.

The only SN Ib-norm  that satisfied our selection cuts in \cite{magill2025super} is SN 2018beh, which is reported in \citet{gomez2022luminous} to be a ``luminous supernovae (LSNe)'', a new class characterized by ``spectra that closely resemble those of SLSNe.'' Thus, our failure to classify the spectra from this object does not reflect a failure mode of \abcsn, but rather label noise and the inclusion of a new class that we had not trained on.

The new SNID sample does not include normal SN Ia since it is designed to enhance the SNID library with more complete subtype samples. Thus, we additionally selected 14 spectra from five nearby, well-studied, bona-fide normal SN Ia's: SN 2014J ($z=0.000677$, \citealt{2014CBET.3792....1F}), SN 2015F ($z=0.0049$, \citealt{2015AAN...515....1W}), SN 2019yvq ($z=0.00908$, \citealt{2019TNSTR2720....1I}), SN 2020nlb (in M85 at a distance of 15.8 Mpc, \citealt{2020TNSAN.126....1T}), and SN 2021aefx ($z=0.005$, \citealt{tartaglia2018early}).  We use at least two spectra for each, preprocessed them with a \texttt{Python} version of the SNID preprocessing tools\footnote{\url{https://github.com/fedhere/snidpy/tree/main}} as described in \autoref{sec:data:prep}. Since the SN are all nearby targets, we did not correct redshift. The phase ranges between -10 and 40 days of maximum date. 
Of the 14 SN Ia spectra, 12 were classified correctly as Ia-norm, two as Ia-91T. Both spectra classified as SN Ia-91T-like are pre-maximum (sn2020nlb, -7 days before maximum, sn2021aefx -1 day). However, we note that the other spectra for the same objects were correctly classified SN Ia-norm, and, in spite of our effort to find bona-fide SN Ia's, both SNe are reported to be somewhat unusual: e.g. SN 2020nlb is described as ``subluminous SN Ia with deep CSM'' (\citealt{2019cxo..prop.5790S}, see also \citealt{2025arXiv251200555I,2024A&A...685A.135W}) and SN 2021aefx showed an early blue bump in photometric observations that remained challenging to model (e.g. \citealt{2022ApJ...933L..45H});  \citealt{2023ApJ...959..132N} reports
``Our spectrum of SN 2021aefx from the epoch of B-band
maximum confirms its intermediate classification
between the CN and BL subtypes of normal Type Ia
SNe, although its near-peak light curves are slow-declining and luminous for this classification, resembling 91T-like peculiar events``. The relative prediction probability for these two events are (Ia-norm, Ia-91T) = (37\%, 60\%) for and  (Ia-norm, Ia-91T) = (15\%, 83\%) for SNe 2020nlb and 2021aefx, respectively. Thus, while based on the extensive studies of these SNe the correct classification is SN Ia-norm, some degree of label confusion is justified by the nature of the objects themselves, and the classification within the broad Ia family is unambiguous. 

These tests overall confirms the transferability of our model to unseen data for all subtypes. The detailed predictions are reported in \autoref{appendix:qub}.

\begin{table}
    \centering
    \begin{tabular}{c|cccc}
        & Precision & Recall & Spectra \\
        \hline
        \hline
        Ia-91T (10) & 1.00 & 0.92  & 24 \\
        Ia-91bg (5) & 1.00 & 0.58  & 12 \\
        Iax (1) & 1.00 &  1.00 & 4 \\
        Ibn (4) & 0.88 & 0.78 & 9 \\
        IIb (9)& 0.59 & 0.84 & 19 \\
        Ic-norm (10) & 0.74 &  0.71 & 24 \\
        Ic-broad (5) & 0.76 & 0.81  & 16 \\
        IIP (43) & 0.91 & 0.84 & 83 \\
        \hline
        macro avg & 0.69 & 0.65 & 194\\
        weighted avg & 0.80 & 0.82 & 194 & \\
    \end{tabular}
    \caption{Performance on data from \citet{magill2025super} unseen in training and comprised of 194 spectra (89 SN). For each subtype, the number of SN is reported in parentheses, and the number of spectra in a dedicated column. The detailed predictions are reported in \autoref{appendix:qub}\label{tab:superfitperformance}}
    
\end{table}




Since we did not train our model with the goal of maximizing the purity of a SN Ia sample, as one would do for a classifier primarily aimed for use in cosmology, we cannot directly compare the performance of \abcsn{} to \sniascore{}. However, we note that from Figure 4 of \citet{fremling_sniascore_2021} at a purity comparable to that of \abcsn{} (5\%) their model accuracy appears to be 92\% compared to our 88\%. Nonetheless, our model can classify 10 subtypes of SNe and is therefore of broader applicability than \sniascore{}.

Harder yet is a comparison with \ccsnscore{} since the model architecture is markedly different: it is a multi-modal (incorporating both lightcurves and spectra) hierarchical model, and their classes are, in some cases, defined differently than ours. Nonetheless, based on their reported performance on subtypes (Section 5.2 and Figure 6 in \citealt{sharma_ccsnscore_2025}) we outperform them on all classes in common: their spectra-only classification accuracy for types (IIb, Ib, Ic, Ic-bl) is (0.15, 0.44, 0.16,  0.50) compared to our (0.70, 0.82, 0.82, 0.82), and their accuracy only improves for IIb when other data modalities (photometry) are included (0.55, 0.36, 0.14, 0.17). We caution the reader, however, that comparison between models should only be done on shared, benchmark datasets (which motivated our work in re-training DASH), so the  comparisons with \sniascore{} and \ccsnscore{} should be treated cautiously.

\section{Conclusion and Future Work} \label{sec:conclusion}
We developed a new, attention based model for the automated classification of supernovae (SNe) spectra into 10 subtypes, including subtypes of thermonuclear, core collapse and stripped envelope supernovae, and we compared its performance with the popular neural network based classifier \dash{} \citep{muthukrishna_dash_2019}.

Our model does require the spectra to be preprocessed through the \snid{} preprocessing steps, including continuum removal, telluric line removal, de-redshifting, zero-padding, and additionally \abcsn{} needs to be served spectra at resolution $R=100$. Alternatively, modifications of the input layer shape and fine tuning would enable other resolutions. This is a potential limitation of its applicability, especially if spectra at high redshift are to be classified, as key features may shift inside the padding region. However, this setup allowed us to directly test and compare different architectures on a homogeneous dataset. Conversely, some relatively small future upgrades to our architecture may include embedding wavelength and flux together before applying positional encoding. While this is not currently necessary since our spectra are evenly sampled, this may improve classification in a broader range of redshifts.

Our model was trained on a dataset of \nSpec{} spectra from \nSN{} SNe, mostly composed of \snid{} template spectra. We lower the resolution of each spectrum from $R = 738$ to $R = 100$ to enable the development and use of an attention-based neural network, which is memory-intensive. In \autoref{sec:data:prep:lowres} and \autoref{fig:performance_vs_R} we demonstrate that lowering the resolution of spectral data does not negatively impact model performance.

To enable a direct and fair comparison and to establish our domain benchmark, we retrained \dash{} to address two issues: overfitting and the contamination of spectra from the same SN in both training and test data (see \autoref{sec:retraindash}). We introduced an early stopping callback to prevent overfitting and rigorously enforced that all spectra from the same SN are only in the training or in the test set.

Special attention is given to the training of \abcsn{}, including ensuring that in train-test split all spectra from a given SN are placed in either the training or the test sets and applying data augmentation to the training set in order to balance the dataset and reduce model bias towards the most numerous classes (see \autoref{sec:data}).

We explored a variety of architectural choices (described in detail in \autoref{sec:methods:abcsn}) and we made our final choice with the goal of delivering a model that reaches high performance with limited overfitting to ensure transferability. Our final model is inspired by the Transformer \citep{vaswani_attention_2017}. It includes a trainable positional encoding (\autoref{sec:methods:abcsn:positionalencoding}), the multi-head attention mechanism as well as a generative, pretraining stage on masked spectra (see \autoref{sec:methods:pretraining})

\abcsn{} improves upon \dash{} as a spectroscopic new-SN classifier, providing higher classification accuracy for 10 SN classes. We measure the performance of \dash{} with the macro F1-score, which accounts for dataset imbalance. \abcsn{} attains a macro F1-score of $82.45\%$ while \dash{}'s macro F1-score retrained on the same dataset reaches $58.86\%$. A detailed performance assessment is provided in \autoref{sec:results} and visualized in \autoref{fig:THE_cm} and \autoref{appendix:cm}. The model recall exceeds 75\% on all classes, the model precision falls below 75\% only for SN Ia-91T ad SN Ic. SN~Ia-91T are contaminated by closely related subtypes of SN Ia (91T, 91bg, and Iax) and of SN Ic by stripped envelope SNe Ib. {The model transferability is confirmed on unseen data as reported in \autoref{sec:results}, \autoref{tab:superfitperformance}, and \autoref{appendix:qub}}. 

While our model is not optimized for cosmology, the purity of the SN Type Ia-norm classification is $95\%$. Since we are releasing this as an open-source software package, the user could easily retrain the model to maximize Ia-norm purity.

\snid{} is still the ubiquitous method for spectroscopic SN classification, and in fact we used \snid{} labels as our ground truth in training. However, AI-based methods offer greater automation options and require limited computational resources once trained, making them well-suited to the era of data-intensive astronomy.

{In this work, we improved upon existing classifiers. In doing so we have tested thousands of different models and we found that our performance is only marginally sensitive to the variety of preprocessing choices we attempted as well as disparate architectural choices. While there remain limitations (\eg{} the limited wavelength range we used), we suspect it is going to be extremely hard to achieve a substantially higher performance on this dataset with improved architectural choices due to label noise, imbalance, and a classification schema based on an assumed rigid taxonomy. }
{Thus, we identify two particularly promising research directions: the curation of a large dataset for spectral classification is critical to effectively benchmark and fairly compare different models. We note some recent developments in this field \citep{milligan2025testing}, yet this work is specific to a future survey and a comprehensive, holistic dataset that enables model performance comparison is still needed. The effectiveness of different architectural choices made in different models will only then become more clear. Second, }
any SN is driven by continuous variables like progenitor mass, progenitor metallicity, circumstellar material composition, etc., and yet we perform classification of SNe into discrete subtypes. Although there do exist some known discrete categories of SNe (\eg{} thermonuclear vs. core collapse), the practice of classifying SNe into rigid subtypes is questionable. Alternative approaches \citep[\eg][]{ williamson2019optimal, kou2020new, s_de_souza_graph-based_2023} and hierarchical classification schemes \citep{villar2023hierarchical, 2025arXiv250101496S}, are being developed and may well result in more meaningful solutions. Nonetheless, classification remains an important and widespread practice as it enables the planning of follow-up resources and the construction of samples that support studies into different SN physical processes. Furthermore, neural network-based classifiers naturally provide a probabilistic classification, which allows users to redefine spectral labels.

\thesis{However, to blame any lack of \abcsn{} performance on the data without acknowledging the limitations of the authors would be hubristic. We certainly hope that there is no label noise and that breakthroughs in SN classification are forthcoming, but in the absence of those breakthroughs and in the aftermath of our significant model testing, we are forced to conclude that label noise may play a significant role in limiting classification performance.}

We further acknowledge that \snid{} and \dash{} provide not only class labels but also redshift and phase estimates. {We applied augmentations specific to increase robustness to incorrect redshift corrections and we further note that recent work \citep{xu2025applecider} shows that convolutional models for spectroscopic classification of astronomical sources are robust to redshift translations. However,} in this work, we only focus on SN subtype classification in order to provide an extensive exploration of neural network solutions and an effective tool for the community at the incipit of the \lsst{} era. We expect the most common application of \abcsn{} will be to classify newly discovered SNe, and we leave it to future work to extend \abcsn{} to those additional tasks. \abcsn{} could also be retrained on a larger dataset that includes non-SN transients; we leave this task as future work as well.

This work also demonstrates that SN classification with low-resolution spectra is perfectly viable. This is an important result since in the era of \lsst{} when spectroscopic resources are extremely limited compared to the wealth of transient discoveries. In separate work, we will provide guidelines on the spectral resolution and signal-to-noise-ratio (SNR) necessary for different transient classification tasks. For the moment, \abcsn{} is designed to classify spectra at the resolution of the \sedmachine{} \citep{blagorodnova_sed_2018}, $R=100$, and we provide methods to lower higher resolution spectra to $R=100$ for the application of \abcsn{}.

\abcsn{} is an open source tool\footnote{https://github.com/FoxFortino/ABC-SN}, we provide a pretrained model ready for application\footnote{https://zenodo.org/records/16620817}, together with tools for preprocessing spectra, and we provide the code base and hope the astrophysical community will continue to refine \abcsn{} for additional applications.

\begin{acknowledgments}
    WFF and FBB acknowledge support from the National Science Foundation Award No. AST-2108841.
    
    The authors acknowledge the support of the Vera C. Rubin Legacy Survey of Space and Time Science Collaborations and particularly of the Transient and Variable Star Science Collaboration (TVS SC), which provided opportunities for collaboration and exchange of ideas and knowledge.

    The authors are grateful for thoughtful and helpful reviews from the anonymous referee. We believe their comments and suggestions truly strengthened our work.
    
    Special thanks are also given to the penguins, polar bears, and all else who live in cold climes. Polar habitats are degraded by heat generated from our computing tasks; we hope their unwilling sacrifice will not be in vain.
\end{acknowledgments}

\begin{table*}
    \centering
    \small
    \begin{center}
    \begin{tabular}{l|lll}
    \multicolumn{3}{c}{} \\
    \multicolumn{3}{c}{\textbf{Preprocessing}} \\
    \hline 
    Full wavelength range & $2500$ \AA{} \textendash{} $10,000$ \AA{} & Wavelength range of the input data.\\
    Non-zero wavelength range & $4500$ \AA{} \textendash{} $7000$ \AA{} & Flux is padded to zero outside of this range.\\
    Supernova phase range & [$-20$, $50$] & Days since estimated to $V$ band maximum. \\
    Peak-to-peak range & $[0.1, 100]$ & Spectra with a larger range are removed from the dataset.\\
    Training set fraction & 50\%  \\
    \multicolumn{3}{c}{} \\
    \multicolumn{3}{c}{\textbf{Data Augmentation}} \\
    \hline 
    Injected noise  & $\sim\mathcal{N}(0, 0.01)$ & Noise distribution.\\
    Spectrum shifting & $N_\mathrm{shift} \in [-5, 5]$  & Shift spectrum by $N_\mathrm{shift}$ wavelength bins drawn from a uniform (integer) \\
    &&distribution in this range.\\
    Number of injected spikes & $N_\mathrm{spikes} \in [0, 4]$ & $N_\mathrm{spikes}$ drawn from a uniform (integer) distribution in this range.\\
    Injected spike sign & $p(+) = 0.8$& Probability of spike being in emission ($+$), otherwise absorption. \\
    Spike amplitude & $2 \sigma_{i}$ & Where $\sigma_{i}$ is the standard deviation of the spectrum being augmented. \\
    
    \multicolumn{3}{c}{} \\
    \multicolumn{3}{c}{\textbf{Pretraining}} \\
    \hline 
    Masking fraction & 15\% & Amount of each spectrum to mask for pretraining. \\
    Mask value & 0 & Set the masked values to this value for pretraining \\
    Batch size & 64 & Batch size of the pretraining network. \\
    \multicolumn{2}{c}{\hspace{-50mm}{\it Early stopping}} \\
    \hspace{3mm}Minimum improvement & $\Delta L=0.0005$ & Minimum improvement threshold in test MSE for early stopping. \\ 
    \hspace{3mm}Patience & $P_e=25$ & Training halted after $P_e$ epochs with test MSE improvement $<\Delta L$.  \\ 
    Initial learning rate ($lr$)& $lr_\mathrm{start}=10^{-4}$ \\
    \multicolumn{2}{c}{\hspace{-12mm}{\it Reduce Learning Rate on Plateau (RLRP)}} \\
    \hspace{3mm}Minimum improvement & $\Delta L_\mathrm{RLRP}=0.0005$ & Minimum improvement in test MSE for $lr$ updates.\\
    \hspace{3mm}Patience & $P_\mathrm{RLRP}=10$ & $lr$ reduced after $P_\mathrm{RLRP}$ epochs with improvement $<\Delta L_\mathrm{RLRP}$. \\
    \hspace{3mm}Factor & 0.5 & Reduce $lr$ by this factor. \\
    \hspace{3mm}Minimum $lr$ & $lr_\mathrm{min}=10^{-7}$ & RLRP will not reduce $lr$ below this threshold. \\

    \multicolumn{3}{c}{} \\
    \multicolumn{3}{c}{\textbf{\abcsn{} Training}} \\
    \hline 
    Batch size & 64 & Batch size of \abcsn{} during training. \\
    \multicolumn{2}{c}{\hspace{-50mm}{\it Early stopping}} \\
    \hspace{3mm}Minimum improvement & $\Delta F1 = 0.005$ & Minimum improvement threshold in test F1. \\
    \hspace{3mm}Patience & $P_e=25 $& Training halted after $P_e$ epochs with test F1 improvement $<\Delta F1$. \\
    Initial learning rate ($lr$) & $lr_\mathrm{start}=10^{-5}$ \\
    \multicolumn{2}{c}{\hspace{-12mm}{\it Reduce Learning Rate on Plateau (RLRP)}} \\
    \hspace{3mm}Minimum improvement & $\Delta L_\mathrm{RLRP}=0.0005$ & Minimum improvement in test F1 for $lr$ updates.\\
    \hspace{3mm}Patience & $P_\mathrm{RLRP}=10$ & $lr$ reduced after $P_\mathrm{RLRP}$ epochs with improvement $<\Delta L_\mathrm{RLRP}.$ \\
    \hspace{3mm}Factor & 0.5 & Reduce $lr$ rate by this factor. \\
    \hspace{3mm}Minimum $lr$ & $lr_\mathrm{min}=10^{-7}$ & RLRP will not reduce $lr$ below this threshold. \\
    \label{tab:hyperparams}
    \end{tabular}
    
    \caption{A table of hyperparameter choices for \abcsn{}. {\it Preprocessing} describes choices in data cleaning and culling (\autoref{sec:data:prep}). {\it Data Augmentation} describes how we duplicated and modified spectra in order to balance our dataset and combat overfitting (\autoref{sec:data:prep:augmentation}). {\it Pretraining} hyperparameters account for the training scheme we used for the masked spectra regression model (\autoref{sec:methods:pretraining}). {\it \abcsn{} Training} hyperparameters detail the training scheme for the predictive model (\autoref{sec:methods:abcsn}).}
    \label{table:hyperparams}
    \end{center}
\end{table*}

\appendix

\section{ABC-SN Training hyperparameters and data preparation choices} \label{appendix:hyperparam}

\abcsn{} training hyperparameters and data preparation choices are reported in \autoref{table:hyperparams}.

\section{Complete performance assessment of \abcsn{}} \label{appendix:cm}

\begin{figure*}[h!]
    \includegraphics[width=0.85\textwidth]{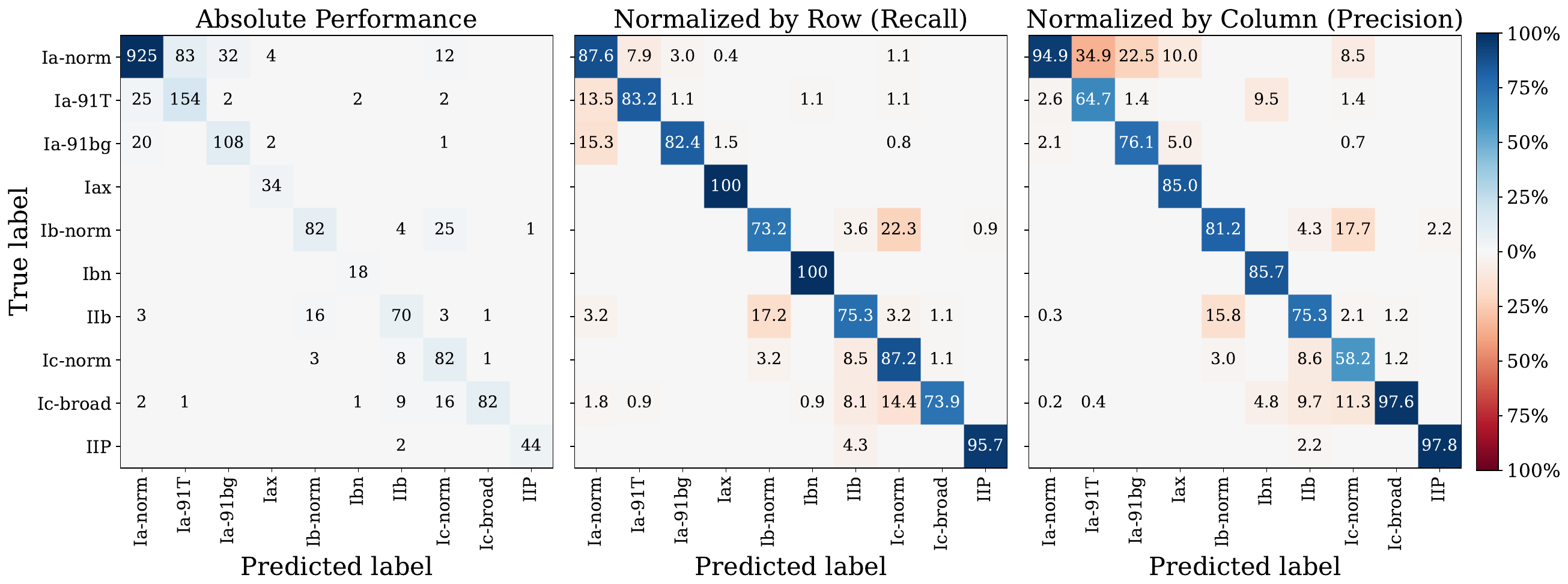}
    \centering
    \caption{Confusion matrix for the release version of \abcsn{}. In the first matrix, the data is not normalized and the numbers and colors reflect the absolute number of SN in each bin. The second matrix is normalized by row (\ie{} recall/completeness). The third matrix is normalized by column (\ie{} precision/purity).}
    \label{fig:abcsn_triple_cm}
\end{figure*}

\autoref{fig:abcsn_triple_cm} shows a holistic view of \abcsn{} performance with three test set confusion matrices. The first shows the absolute performance of \abcsn{}, providing raw values for the number of classified spectra. The second confusion matrix shows the same data but normalized by the number of true labels for each class. The percentages along the diagonal represent the completeness of \abcsn{} for each class, answering the question, ``What fraction of Class X was correctly classified?'' The third confusion matrix shows the same data again but normalized by the number of predicted labels for each class. The percentages along the diagonal represent the purity of \abcsn{} predictions, answering the question, ``What fraction of spectra labeled as Class X were indeed Class X?''

\begin{figure*}[h!]
    \includegraphics[width=0.85\textwidth]{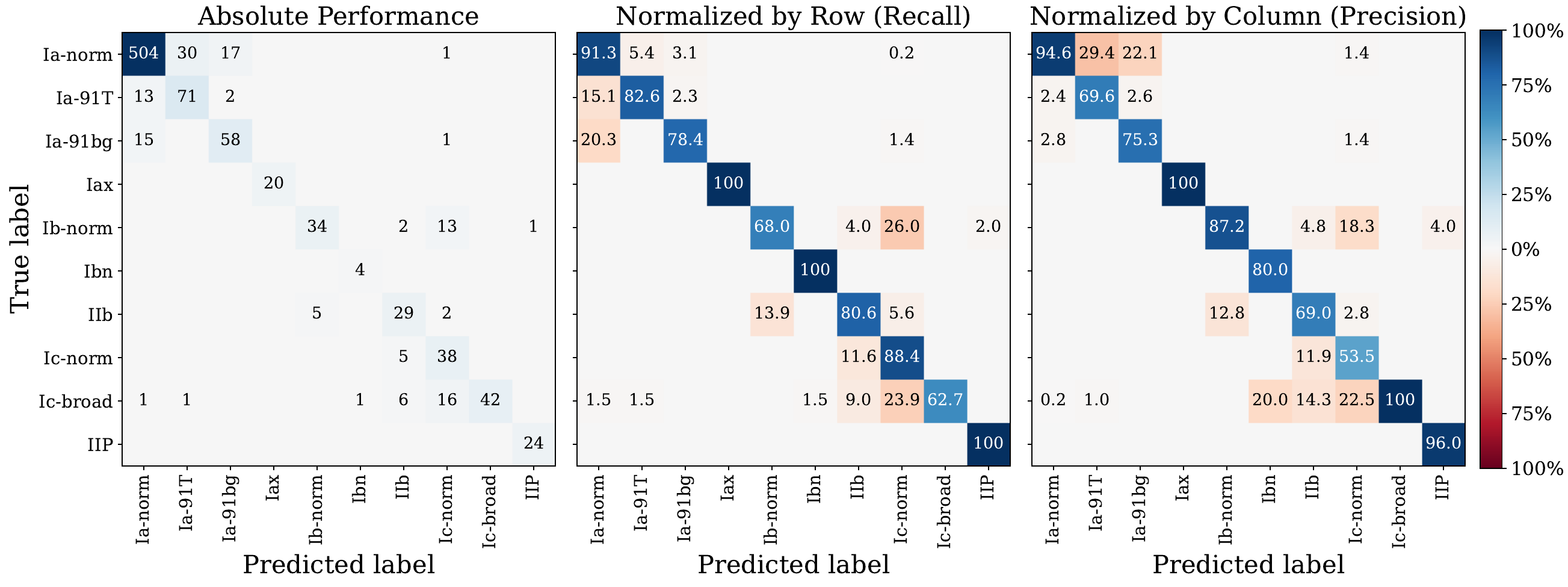}
    \centering
    \caption{{Same as \autoref{fig:abcsn_triple_cm} for spectra in the epoch range -10 to 10 days of maximum light only.}}
    \label{fig:abcsn_triple_cm_neapeak}
\end{figure*}

\begin{figure*}[h!]
    \includegraphics[width=0.85\textwidth]{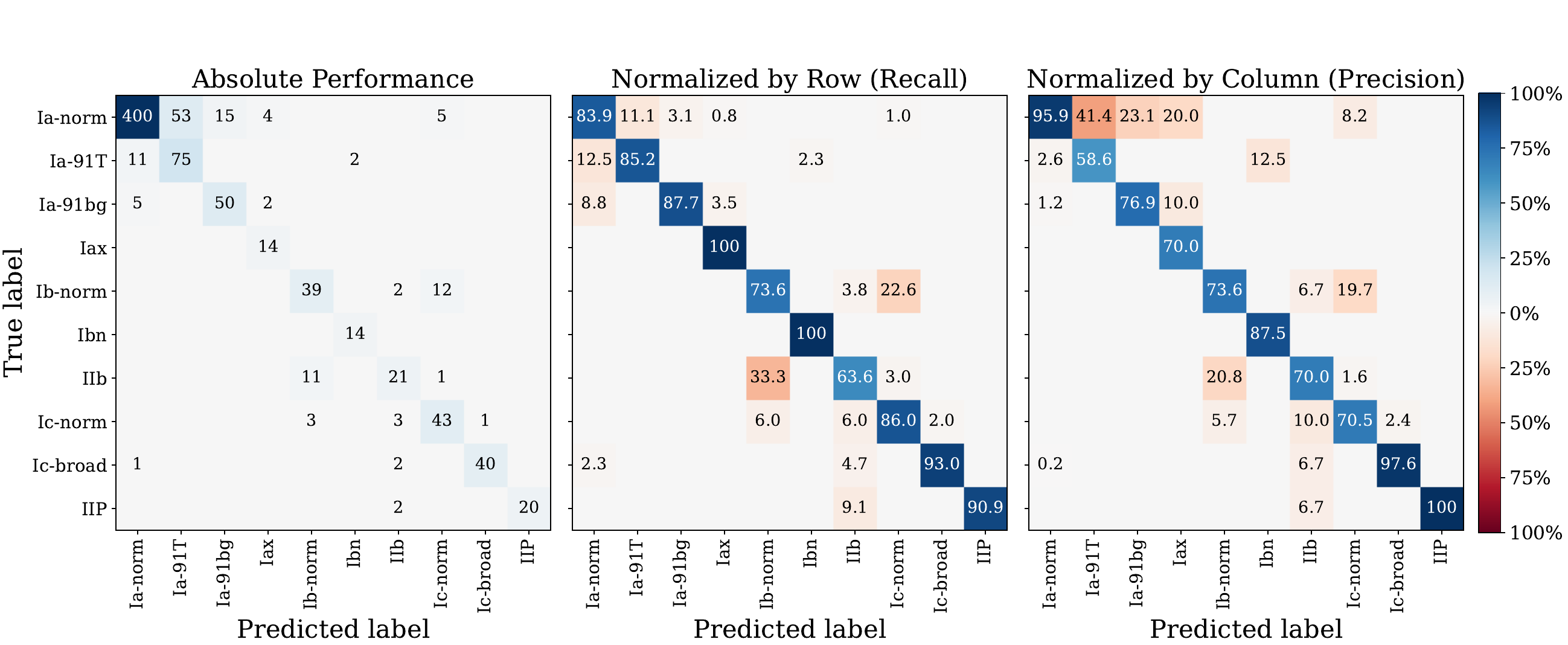}
    \centering
    \caption{{Same as \autoref{fig:abcsn_triple_cm} for spectra post 10 days of maximum light only.}}
    \label{fig:abcsn_triple_cm_postpeak}
\end{figure*}

{Confusion metrics are also shown for the subset of test spectra near peak (-10 to 10 days of maximum light) and post peak (10 to 50 days) in \autoref{fig:abcsn_triple_cm_neapeak} and \autoref{fig:abcsn_triple_cm_postpeak}, respectively.}

\section{Performance on Super-SNID Templates} \label{appendix:qub}

We tested the performance of our model on unseen data and reported performance metrics in \autoref{sec:results} and \autoref{tab:superfitperformance}. The data is collected from \citet{magill2025super}: 194 spectra (89 SNe) in the phase range -10 to 20 days since maximum brightness. The detailed classification results are shown in \autoref{tab:superfit} and discussed in \autoref{sec:results}. Classifications inconsistent with the provided SNID label are marked in bold in \autoref{tab:superfit}: \abcsn{} predicts different classes than the label for 21 spectra (of 13 SNe), but  6 of those SNe (13 spectra) have been shown to be unusual or misclassified in the literature (see footnotes 12-17). In addition, predictions for 14 SN Ia spectra selected as described in \autoref{sec:results} are included at the end of \autoref{tab:superfitperformance}. The sample was assembled by (1) identifying in the literature well-studied SN Ia with unambiguous classification discovered after 2014 to ensure they would not be included in our initial dataset ($\sim10$ SNe); (2) searching for optical spectra in WISeREP \citep{2012PASP..124..668Y} (3) ensuring the classification on WISeREP is unambiguously consistent with SN Ia-norm (this surprisingly removed several SNe); (4) selecting a few spectra per SN in the range (-10---40) days from peak; (5) ensuring each spectrum has good coverage in the range 2,500-10,000~\AA. The final sample includes the following spectra: 
\begin{itemize}
    \item SN 2015F
    \begin{itemize}
        \item 
        SN2015F\_2015-03-19\_01-12-24\_ESO-NTT\_EFOSC2-NTT\_PESSTO\_SSDR1-4.csv;
        \item SN2015F\_2015-03-20\_23-52-27\_ESO-NTT\_EFOSC2-NTT\_PESSTO\_SSDR1-4.csv;
        \item SN2015F\_2015-03-28\_02-40-05\_ESO-NTT\_EFOSC2-NTT\_PESSTO\_SSDR1-4.csv;
        \item SN2015F\_2015-04-11\_03-01-13\_ESO-NTT\_EFOSC2-NTT\_PESSTO\_SSDR1-4.csv.
    \end{itemize}
    \item SN 2014J
    \begin{itemize}
        \item SN2014J\_2014-01-31\_00-00-00\_HCT-2m\_HFOSC\_None.dat;
        \item SN2014J\_2014-03-07\_03-15-50\_Lick-3m\_KAST\_UCB-SNDB.flm;
        \item SN2014J\_2014-03-13\_00-00-00\_HCT-2m\_HFOSC\_None.dat.
   \end{itemize}
   \item SN 2019yvq
   \begin{itemize}
        \item 2019yvq\_2020-01-15\_09-09-01\_P60\_SEDM\_ZTF.ascii;
        \item 2019yvq\_2020-01-24\_14-44-38\_Keck1\_LRIS\_ZTF.ascii;
        \item 2020nlb\_2020-07-06.0\_LCO2m\_Spectral\_None.txt.
   \end{itemize}
   \item 2020nlb
   \begin{itemize}
        \item 2020nlb\_2020-07-19\_LCO2m\_Spectral\_None.txt;
        \item 2020nlb\_2020-07-27.0\_Lick-3m\_KAST\_None.csv.
   \end{itemize}
   \item 2021aefx
   \begin{itemize}
        \item 2021aefx\_2021-11-28.12\_Gemini-S\_GMOS-S\_None.ascii;
        \item 2021aefx\_2021-12-12\_23-48-32.500\_SALT\_RSS\_GSP.txt.
    \end{itemize}
\end{itemize}

\startlongtable
\begin{deluxetable}{lrll}
\tablecaption{\abcsn{} classification of transients unseen in training and test sets from \citep{magill2025super} and selected SN Ia's.}

\tablehead{\colhead{SN} & \colhead{Phase} & \colhead{SNID Label} & \colhead{ABC-SN Label}}

\startdata
sn2012dy & -4.0 & IIb & IIb \\
sn2012dy & -2.0 & IIb & IIb \\
sn2012dy & 7.0 & IIb & IIb \\
sn2012dy & 13.0 & IIb & IIb \\
sn2012dy & 15.0 & IIb & IIb \\
sn2012ec & 7.0 & IIP & IIP \\
sn2012ec & 15.0 & IIP & IIP \\
LSQ12dwl & -6.0 & Ic-norm & Ic-norm \\
LSQ12dwl & -4.0 & Ic-norm & Ic-norm \\
LSQ12dwl & 3.0 & Ic-norm & Ic-broad \\
LSQ12dwl & 10.0 & Ic-norm & Ic-broad \\
LSQ12gdj & -8.0 & Ia-91T & Ia-91T \\
LSQ12gdj & -6.0 & Ia-91T & Ia-91T \\
LSQ12gdj & 2.0 & Ia-91T & Ia-91T \\
LSQ12gdj & 15.0 & Ia-91T & Ia-91T \\
sn2013k & 7.0 & IIP & IIP \\
sn2013k & 13.0 & IIP & IIP \\
sn2013k & 16.0 & IIP & IIP \\
\textbf{sn2013u} & 14.0 & Ia-91T & Ia-norm \\
sn2013ai & 8.0 & IIP & IIP \\
sn2013ai & 10.0 & IIP & IIP \\
sn2013ai & 16.0 & IIP & IIb \\
sn2013ai & 18.0 & IIP & IIb \\
sn2013ak & 18.0 & IIb & IIb \\
sn2013bb & -6.0 & IIb & IIb \\
sn2013bb & 1.0 & IIb & IIb \\
sn2013bb & 7.0 & IIb & IIb \\
sn2013fq & 18.0 & IIb & IIb \\
sn2013fs & -3.0 & IIP & IIP \\
sn2014l & -5.0 & Ic-norm & Ic-norm \\
sn2014l & 0.0 & Ic-norm & Ic-norm \\
sn2014l & 2.0 & Ic-norm & Ic-norm \\
sn2014cx & -5.0 & IIP & IIP \\
sn2014cx & 2.0 & IIP & IIP \\
sn2014cx & 9.0 & IIP & IIP \\
sn2014dq & 1.0 & IIP & IIP \\
sn2014dq & 2.0 & IIP & IIP \\
sn2014eg & -8.0 & Ia-91T & Ia-91T \\
sn2014eg & -6.0 & Ia-91T & Ia-91T \\
sn2014eg & -2.0 & Ia-91T & Ia-91T \\
sn2014eg & 0.0 & Ia-91T & Ia-91T \\
sn2014eg & 13.0 & Ia-91T & Ia-91T \\
sn2014eg & 20.0 & Ia-91T & Ia-91T \\
CSS140421 & 5.0 & Ibn & Ibn \\
CSS140421 & 6.0 & Ibn & Ibn \\
CSS140421 & 13.0 & Ibn & Ibn \\
CSS140421 & 17.0 & Ibn & Ibn \\
ASASSN-14ha & 5.0 & IIP & IIP \\
ASASSN-14ha & 5.0 & IIP & IIP \\
ASASSN-14ha & 6.0 & IIP & IIP \\
ASASSN-14ha & 12.0 & IIP & IIP \\
ASASSN-14ha & 14.0 & IIP & IIP \\
ASASSN-14ha & 19.0 & IIP & IIP \\
sn2015H & 6.0 & Iax & Iax \\
sn2015H & 16.0 & Iax & Iax \\
sn2015H & 10.0 & Iax & Iax \\
sn2015H & 3.0 & Iax & Iax \\
sn2015ay & 0.0 & IIP & IIb \\
sn2015bm & 4.0 & IIP & IIP \\
sn2015bm & 5.0 & IIP & IIb \\
sn2015bm & 20.0 & IIP & IIP \\
sn2015bo & -10.0 & Ia-91bg & Ia-norm \\
sn2015bo & -6.0 & Ia-91bg & Ia-norm \\
sn2015bo & 17.0 & Ia-91bg & Ia-norm \\
sn2015bo & 17.0 & Ia-91bg & Ia-norm \\
sn2015bs & 1.0 & IIP & IIb \\
sn2015bs & 15.0 & IIP & IIP \\
sn2015ch & 0.0 & IIP & IIb \\
sn2015ch & 1.0 & IIP & IIP \\
sn2015ch & 4.0 & IIP & IIP \\
sn2015ch & 16.0 & IIP & IIP \\
\textbf{PSNJ15053007} & 0.0 & Ia-91bg & Ia-norm \\
PSNJ15053007 & 8.0 & Ia-91bg & Ia-91bg \\
PS15cww & 6.0 & IIP & IIP \\
PS15cww & 18.0 & IIP & IIP \\
\textbf{sn2016p}\footnote{The classification of this supernova is contested in the literature \citep{gangopadhyay_optical_2020, prentice_investigating_2019,finneran_velocity_2025}.} & -1.0 & Ic-broad & Ic-norm \\
\textbf{sn2016p} & 7.0 & Ic-broad & IIP \\
\textbf{sn2016p} & 18.0 & Ic-broad & IIP \\
sn2016x & -2.0 & IIP & IIP \\
sn2016x & 7.0 & IIP & IIP \\
sn2016x & 14.0 & IIP & IIP \\
\textbf{sn2016adj}\footnote{The classification of this supernova appears contested, showing unusual carbon features \citep{stritzinger_carbon-rich_2024}.} & -2.0 & IIb & IIP \\
\textbf{sn2016adj} & 2.0 & IIb & Ib-norm \\
\textbf{sn2016adj} & 8.0 & IIb & Ic-broad \\
sn2016aiy & -6.0 & IIP & IIP \\
sn2016aiy & 8.0 & IIP & IIP \\
sn2016aqf & 16.0 & IIP & IIP \\
sn2016blz & 1.0 & IIP & IIP \\
sn2016blz & 4.0 & IIP & IIP \\
sn2016blz & 6.0 & IIP & IIP \\
sn2016dsg & 4.0 & Ic-broad & Ic-broad \\
sn2016dsg & 14.0 & Ic-broad & Ic-broad \\
sn2016egz & 7.0 & IIP & IIP \\
sn2016egz & 7.0 & IIP & IIP \\
sn2016egz & 8.0 & IIP & IIP \\
sn2016enp & 6.0 & IIP & IIP \\
sn2016enp & 7.0 & IIP & IIP \\
sn2016enp & 7.0 & IIP & IIP \\
sn2016gkg & -10.0 & IIb & IIb \\
sn2016gkg & -9.0 & IIb & IIb \\
sn2016gkg & -4.0 & IIb & IIb \\
sn2016gkg & 1.0 & IIb & IIb \\
sn2016gkg & 12.0 & IIb & IIb \\
sn2016iae & -4.0 & Ic-norm & Ic-norm \\
sn2016iae & -2.0 & Ic-norm & Ic-norm \\
sn2016iae & -1.0 & Ic-norm & Ic-norm \\
sn2016iae & 0.0 & Ic-norm & Ic-norm \\
sn2016ije & 4.0 & Ia-91bg & Ia-91bg \\
sn2016ije & 7.0 & Ia-91bg & Ia-91bg \\
sn2016iyd & 1.0 & IIP & IIP \\
sn2016iyd & 2.0 & IIP & IIP \\
sn2016iyd & 3.0 & IIP & IIP \\
sn2016iyd & 10.0 & IIP & IIP \\
sn2016iyd & 18.0 & IIP & IIP \\
\textbf{sn2017abw} & 4.0 & IIP & Ic-norm \\
sn2017abw & 15.0 & IIP & IIP \\
sn2017abw & 16.0 & IIP & IIP \\
sn2017awz & -10.0 & Ia-91T & Ia-91T \\
sn2017awz & -4.0 & Ia-91T & Ia-91T \\
sn2017awz & 5.0 & Ia-91T & Ia-91T \\
sn2017awz & 7.0 & Ia-91T & Ia-91T \\
sn2017dcc & 6.0 & Ic-broad & Ic-broad \\
sn2017dcc & 7.0 & Ic-broad & Ic-broad \\
sn2017dcc & 13.0 & Ic-broad & Ic-broad \\
sn2017dcc & 15.0 & Ic-broad & Ic-broad \\
sn2017dfb & -9.0 & Ia-91T & Ia-91T \\
sn2017dfb & -6.0 & Ia-91T & Ia-91T \\
sn2017dio & -5.0 & Ic-norm & IIP \\
sn2017dio & -4.0 & Ic-norm & IIP \\
sn2017gax & 5.0 & Ic-norm & Ic-norm \\
sn2017gax & 8.0 & Ic-norm & Ic-norm \\
sn2017gax & 14.0 & Ic-norm & Ic-norm \\
sn2017gmr & 7.0 & IIP & IIb \\
\textbf{sn2017hbj\footnote{\citealt{pessi2023broad} reports SN2017hbj belngs to group of ``LSNe II that show peculiarities in their $H\alpha$ profile''}} & 13.0 & IIP & Ic-norm \\
sn2017hbj & 14.0 & IIP & IIb \\
sn2017hrq & -7.0 & IIP & IIP \\
sn2017hrq & -5.0 & IIP & IIP \\
sn2017hrq & 11.0 & IIP & IIP \\
sn2017hyh & -5.0 & IIb & IIb \\
sn2017iuk & -3.0 & Ic-broad & Ic-broad \\
sn2017iuk & -1.0 & Ic-broad & Ic-broad \\
sn2017iuk & 1.0 & Ic-broad & Ic-broad \\
sn2017iuk & 10.0 & Ic-broad & Ic-broad \\
sn2017iuk & 11.0 & Ic-broad & Ic-broad \\
sn2017iuk & 13.0 & Ic-broad & Ic-broad \\
sn2017iuk & 14.0 & Ic-broad & Ic-broad \\
sn2017jei & -10.0 & IIP & IIP \\
sn2017jei & 5.0 & IIP & IIb \\
\textbf{sn2017jei} & 7.0 & IIP & Ibn \\
\textbf{sn2018ec} & -10.0 & Ic-norm & IIP \\
\textbf{sn2018ec} & -5.0 & Ic-norm & IIP \\
sn2018apo & -9.0 & Ia-91T & Ia-91T \\
sn2018apo & 3.0 & Ia-91T & Ia-91T \\
\textbf{sn2018beh}\footnote{See \autoref{sec:results}: this SN appears to belong to a subclass of high luminosity SNe \citep{gomez2022luminous} and not be a normal SN Ib.} & -1.0 & Ib-norm & Ic-norm \\
\textbf{sn2018beh} & 5.0 & Ib-norm & Ic-norm \\
\textbf{sn2018beh} & 16.0 & Ib-norm & Ic-norm \\
sn2018bie & -7.0 & Ia-91T & Ia-91T \\
sn2018bsz & -8.0 & IIP & IIP \\
sn2018bsz & -6.0 & IIP & IIP \\
sn2018emt & 10.0 & IIP & IIP \\
sn2018emt & 16.0 & IIP & IIP \\
sn2018eph & 4.0 & IIP & IIP \\
sn2018eph & 11.0 & IIP & IIP \\
sn2018eph & 16.0 & IIP & IIP \\
sn2018evy & 0.0 & IIP & IIP \\
sn2018evy & 8.0 & IIP & IIP \\
sn2018fus & 7.0 & IIP & IIP \\
sn2018fus & 13.0 & IIP & IIP \\
\textbf{sn2018gjx\footnote{\citet{prentice2020sn} reports this as an unusual low-luminosity transient with three distinct spectroscopic phases, CSM interaction, the appearance of IIb-like features, and interaction with 
a helium-rich CSM.}} & 1.0 & Ibn & Ib-norm \\
\textbf{sn2018gjx} & 16.0 & Ibn & IIb \\
sn2018gsk & 4.0 & Ic-norm & Ic-norm \\
sn2018gsk & 11.0 & Ic-norm & Ic-norm \\
sn2018gsk & 19.0 & Ic-norm & Ic-norm \\
sn2018hjw & 6.0 & Ia-91T & Ia-91T \\
\textbf{sn2018iuq} & 13.0 & IIP & IIb \\
sn2018jkb & -8.0 & IIP & IIP \\
sn2018jkb & -2.0 & IIP & IIP \\
sn2018jmt & -2.0 & Ibn & Ibn \\
sn2018jmt & -1.0 & Ibn & Ibn \\
sn2018jmt & 14.0 & Ibn & Ibn \\
sn2018kpo & 9.0 & IIP & IIP \\
sn2018ldu & 15.0 & IIP & IIP \\
sn2019so & -9.0 & Ia-91bg & Ia-91bg \\
sn2019so & -7.0 & Ia-91bg & Ia-91bg \\
sn2019so & 1.0 & Ia-91bg & Ia-91bg \\
sn2019so & 6.0 & Ia-91bg & Ia-91bg \\
sn2019akg & 12.0 & Ia-91T & Ia-91T \\
\textbf{sn2019ape} & 9.0 & Ic-norm & Ic-broad \\
sn2019asz & 1.0 & IIP & IIP \\
sn2019asz & 18.0 & IIP & IIP \\
sn2019bka & -4.0 & Ia-91T & Ia-91T \\
\textbf{sn2019bka}\footnote{This appears to be an unusual Ia-91T \citep{dimitriadis_ztf_2025}.}& 4.0 & Ia-91T & Ia-norm \\
sn2019bkc & 3.0 & Ic-norm & Ic-norm \\
sn2019bkc & 5.0 & Ic-norm & Ic-norm \\
\hline
sn2014J & -1.6 & Ia-norm & Ia-norm \\
sn2014J & 33.4 & Ia-norm & Ia-norm \\
sn2014J & 39.4 & Ia-norm & Ia-norm \\
sn2015F & -6.5 & Ia-norm & Ia-norm \\
sn2015F & -5.5 & Ia-norm & Ia-norm \\
sn2015F & 2.6 & Ia-norm & Ia-norm \\
sn2015F & 16.6 & Ia-norm & Ia-norm \\
sn2019yvq & 0.2 & Ia-norm & Ia-norm \\
sn2019yvq & 9.2 & Ia-norm & Ia-norm \\
\textbf{sn2020nlb\footnote{But see \citet{2025arXiv251200555I} titled ``Normal or transitional? The evolution and properties of two type Ia supernovae in the Virgo cluster''}} & -7.0 & Ia-norm & Ia-91T \\
sn2020nlb & 6.0 & Ia-norm & Ia-norm \\
sn2020nlb & 14.0 & Ia-norm & Ia-norm \\
\textbf{sn2021aefx\footnote{But see \citet{2023ApJ...959..132N}}} & -0.6 & Ia-norm & Ia-91T \\
sn2021aefx & 13.4 & Ia-norm & Ia-norm \\
\enddata
\label{tab:superfit}
\end{deluxetable}

\bibliography{references,additionalreferences}{}
\bibliographystyle{aasjournal}

\end{document}